# Electric-field tunable spin texture in dual-gated $HfS_2$/graphene van der Waals heterostructures

Haozhe Yang[1,2,*], Zhendong Chi[1], Eoin Dolan[1], Josep Ingla-Aynés[1,3], Beatriz Martín-García[1,4,6], Witold Skowroński[1,5], Marco Gobbi[4,6], Andrey Chuvilin[1,4], Luis E. Hueso[1,4], Fèlix Casanova[1,4,*]

[1] CIC nanoGUNE BRTA, 20018 Donostia-San Sebastian, Basque Country, Spain.

[2] Fert Beijing Institute, MIIT Key Laboratory of Spintronics, School of Integrated Circuit Science and Engineering, Beihang University, China.

[3] Francis Bitter Magnet Laboratory and Plasma Science and Fusion Center, Massachusetts Institute of Technology, Cambridge, MA 02139, USA.

[4] IKERBASQUE, Basque Foundation for Science, 48009 Bilbao, Basque Country, Spain.

[5] AGH University of Krakow, Institute of Electronics, 30-059 Kraków, Poland.

[6] Centro de Física de Materiales (CSIC-EHU), 20018 Donostia-San Sebastian, Basque Country, Spain.

[*]E-mail: haozheyang.buaa@gmail.com; f.casanova@nanogune.eu

## ABSTRACT

Spin–charge interconversion provides the essential link between spin information and electrical signals in spintronic devices. In spin–orbit coupled systems, spin–charge interconversion can contain multiple components associated with different spin-polarization directions, reflecting the underlying spin texture. Tuning this spin texture therefore provides a way to control not only the magnitude of spin–charge interconversion, but also the balance among its different components. Graphene-based van der Waals heterostructures provide an attractive platform for this purpose because interfacial spin–orbit coupling can host multiple spin–charge interconversion channels that couple to different spin-polarization directions. Here, we demonstrate electric-field control of spin texture in a dual-gated $HfS_2$/graphene van der Waals heterostructure, where independent control of carrier density and vertical electric field allows us to disentangle electrostatic doping from electric field-induced spin–orbit effects. Nonlocal spin precession measurements show that near zero vertical electric field, spin–charge interconversion is dominated by a spin Hall component with out-of-plane spins. Increasing the vertical electric field enhances additional in-plane spin components with Rashba–Edelstein-like character, including both parallel and perpendicular to the charge current contributions. These results provide experimental evidence that a vertical electric field can reconfigure the effective spin texture and control the corresponding spin–charge interconversion components, demonstrating electrical tunability of spin–orbit functionalities.

## INTRODUCTION

Spin–charge interconversion is one of the central functionalities in spintronics because it enables the electrical generation and detection of spin information[1,2]. Its efficiency, polarity, and spin selectivity are intimately linked to the underlying spin texture of the electronic states[3–8], making spin texture itself a key property for both fundamental understanding of spin-orbit coupling (SOC) and device operation[9–12]. Representative mechanisms of spin–charge interconversion include the spin Hall effect (SHE) [8–11,13–18] and the Rashba–Edelstein effect (REE) [5,8,12,15,19]. More generally, the form of spin–charge interconversion is constrained by crystal and interfacial symmetry, while structural or interfacial inversion asymmetry can qualitatively reshape the allowed spin polarization and conversion channels, as seen in low-symmetry and chiral systems[20]. An applied electric field therefore provides an especially attractive means of control, as it can continuously modify the inversion-asymmetric electronic environment and thereby tune the spin texture together with the resulting spin–charge interconversion[1,7,8]. Controlling symmetry in this manner offers a direct route toward more unconventional spintronic functionalities.

Among the available materials platforms, graphene-based van der Waals (vdW) heterostructures are especially compelling because graphene combines long-distance spin transport[21,22] and electrostatic tunability with an electronic structure that can inherit strong proximity-induced SOC from an adjacent layered crystal, such as transition-metal dichalcogenide (TMDC)[8–11,18,23–27], monochalcogenide[28], metal oxide[29,30] or topological insulators[31]. In graphene/TMDC heterostructures, this proximity-induced SOC has already manifested in strongly anisotropic spin relaxation[25–27] and spin–charge interconversion[4–6]. Theory and experiment have further shown that the relative alignment between the two layers can reshape the induced spin–orbit fields, with the twist angle directly reconfiguring the spin texture and the associated interconversion response[19,32–36]. In particular, the in-plane spin texture can acquire both tangential and radial components[8,18], which are linked to conventional and unconventional Rashba–Edelstein effects (UREE), respectively. Electrostatic control has also been shown to strongly tune the conversion efficiency itself in proximitized graphene, as demonstrated by gate-tunable SHE and REE[10,11,23,28]. Going beyond efficiency tuning, theory has suggested that an applied electric field can reconfigure the proximity-induced spin texture itself, for example through field-effect spin–valley locking and spin–orbit-valve behavior[19,37,38]. However, experimental studies that directly connect electric-field-controlled spin-texture reconfiguration with its immediate manifestation in spin–charge interconversion remain comparatively limited.

Here we demonstrate electric-field control of spin texture and the resulting spin–charge interconversion in a dual-gated $HfS_2$/graphene vdW heterostructure. In this structure, the upper $HfS_2$ layers oxidize into $HfO_2$, which, together with the $SiO_2$/Si back gate, enables independent control of carrier density and electric field. Using a nonlocal lateral spin-valve geometry and spin precession measurements, we electrically inject spins into proximitized graphene and detect the resulting spin-to-charge conversion (SCC) signal in the nonlocal resistance. At low gate-induced vertical electric fields, SCC in the proximitized graphene is dominated by a spin Hall response associated with a predominantly out-of-plane spin texture. With increasing electric field, the modified inversion asymmetry strengthens Rashba-type contributions and brings out in-plane spin-texture components, leading to both conventional and unconventional Rashba–Edelstein responses. Meanwhile, the spin Hall contribution is progressively suppressed, revealing a clear contrast between the electric-field dependences of the out-of-plane and Rashba-type in-plane spin texture. These results establish electric field as a direct handle for reconfiguring spin texture in a graphene-based heterostructure and point toward voltage-programmable spin–orbit functionalities.

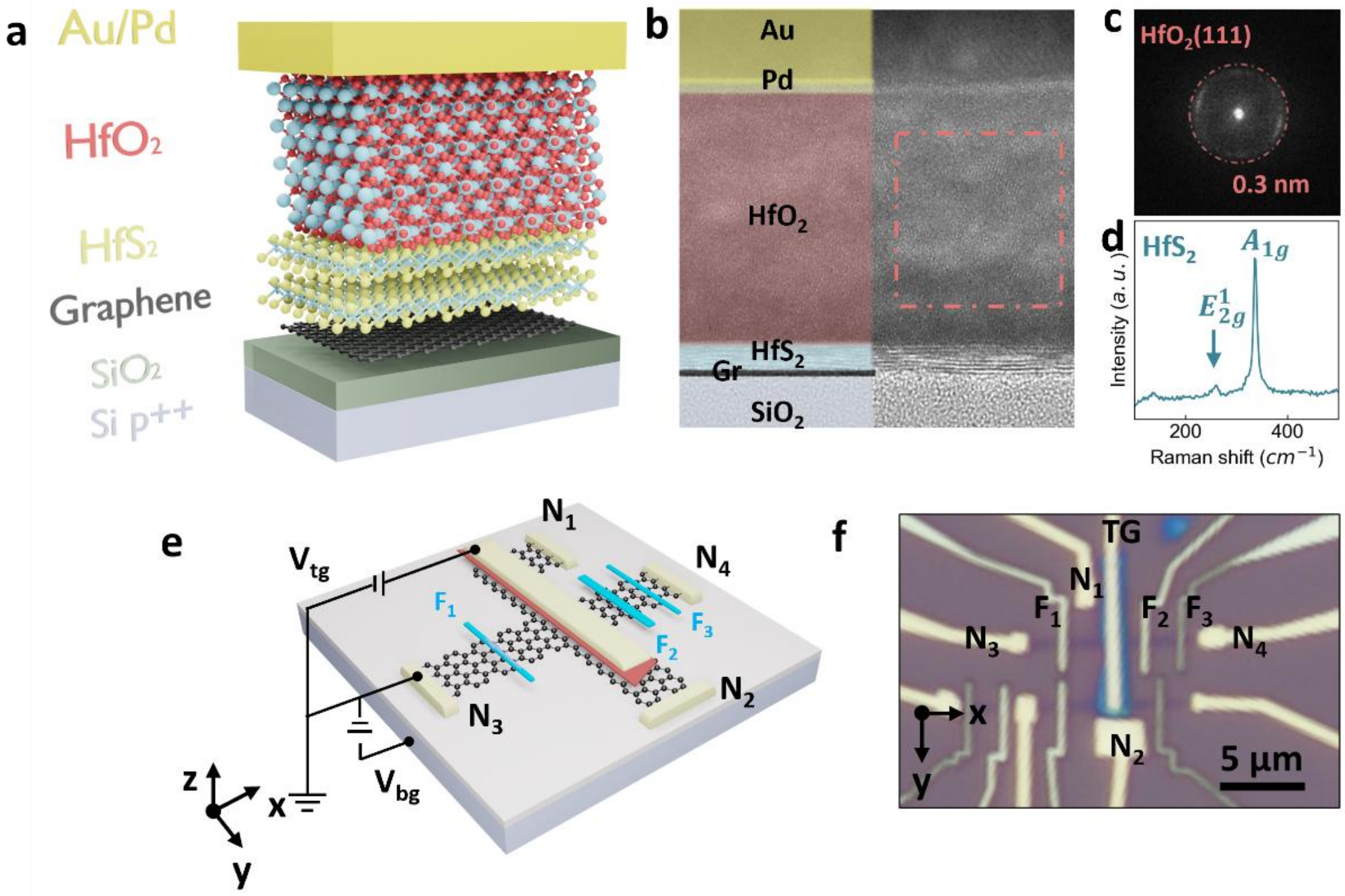

**Fig. 1 Structure and characterization of dual-gated $HfO_2$/$HfS_2$/graphene lateral spin valve devices. a,** Schematic illustration of the vdW heterostructure. An exfoliated $HfS_2$ flake is transferred onto trilayer graphene on a highly doped Si/$SiO_2$ substrate. Subsequent exposure to air partially oxidizes the upper region of $HfS_2$, forming a native oxide layer that serves as the top-gate dielectric. A Pd/Au metal electrode is deposited on top of $HfO_2$ as the local top gate. **b,** Cross-sectional transmission electron microscopy image of the heterostructure stack. The left half is false-colored to highlight the Au/Pd top electrode, $HfO_2$, residual $HfS_2$, graphene and $SiO_2$ layers, while the right half shows the corresponding raw TEM contrast. The remaining $HfS_2$ thickness is approximately 3 nm and the oxide layer thickness is approximately 25 nm. **c,** Local fast Fourier transform analysis taken from the oxidized region marked by the red dashed box in **b**, with the diffraction feature giving an interplanar spacing of approximately 0.30 nm, which corresponds to the $HfO_2$(111) lattice planes. **d,** Raman spectrum measured on an uncovered oxidized $HfO_2$/$HfS_2$ region. The characteristic $E_{2g}^1$ and $A_{1g}$ modes indicate that crystalline $HfS_2$ remains underneath the oxide layer. **e,** Sketch of the dual-gated non-local lateral spin valve device. The graphene channel is patterned into a Hall-bar geometry and contacted by non-magnetic Pd/Au electrodes, labeled $N_1$–$N_4$, and ferromagnetic $TiO_2$/Co/Au tunnel contacts, labeled $F_1$–$F_3$. The top-gate voltage, $V_{tg}$, and back-gate voltage, $V_{bg}$, can be combined to independently control the carrier density and vertical electric field in the $HfS_2$/graphene channel. **f,** Optical micrograph of the device, showing the top-gated $HfO_2$/$HfS_2$/graphene channel, non-magnetic contacts and ferromagnetic electrodes.

## RESULTS AND DISCUSSION

The $HfS_2$/graphene vdW heterostructure was fabricated using a dry-transfer method[39]. $HfS_2$ flakes were mechanically exfoliated onto a PDMS stamp and then transferred onto a trilayer graphene flake on a highly doped Si/$SiO_2$ substrate under an Ar atmosphere. The thickness of trilayer graphene was confirmed by Raman spectroscopy, as shown in Supporting Note 2. $HfS_2$ is a semiconducting TMDC with a reported band gap of about 2 eV[40,41] and is susceptible to oxidation under ambient conditions[42–44]. After transfer, the heterostructure was exposed to air for 30 min, forming a native $HfO_2$ layer on the upper part of the $HfS_2$ flake, as schematically illustrated in Fig. 1a. Because of its large band gap and low conductivity under the gate conditions used here, the residual $HfS_2$ behaves as an insulating layer. Thus, charge and spin transport in the heterostructure occur mainly in the graphene channel. A cross-sectional transmission electron microscopy (TEM) image of the stack is shown in Fig. 1b, with the left half false-colored to identify the different layers. The image reveals a remaining $HfS_2$ thickness of ~3 nm and an oxidized $HfO_2$ layer thickness of ~25 nm. $HfS_2$ oxidation can progressively slow and exhibit self-limiting behavior, while its kinetics and resulting oxide thickness depend on exposure time and temperature[45]. Elemental mapping analysis in Supporting Note 3 shows that sulfur remains detectable in the oxidized region, indicating that the upper $HfS_2$ is only partially oxidized rather than completely sulfur-free. Nevertheless, local fast Fourier transform (FFT) analysis of the selected oxidized region marked by the red dashed box in Fig. 1b, shown in Fig. 1c, gives an interplanar spacing of ~0.30 nm, consistent with the $HfO_2$(111) planes[46]. Therefore, the top dielectric can be regarded as an $HfO_2$-rich $HfO_xS_y$ gate stack, which we refer to as $HfO_2$ for simplicity. Raman spectra were collected from an uncovered oxidized $HfS_2$ region after the electrical measurements, as shown in Fig. 1d. The clear $E_{2g}^1$ and $A_{1g}$ peaks confirm the $HfS_2$ structure[44] remains after oxidation.

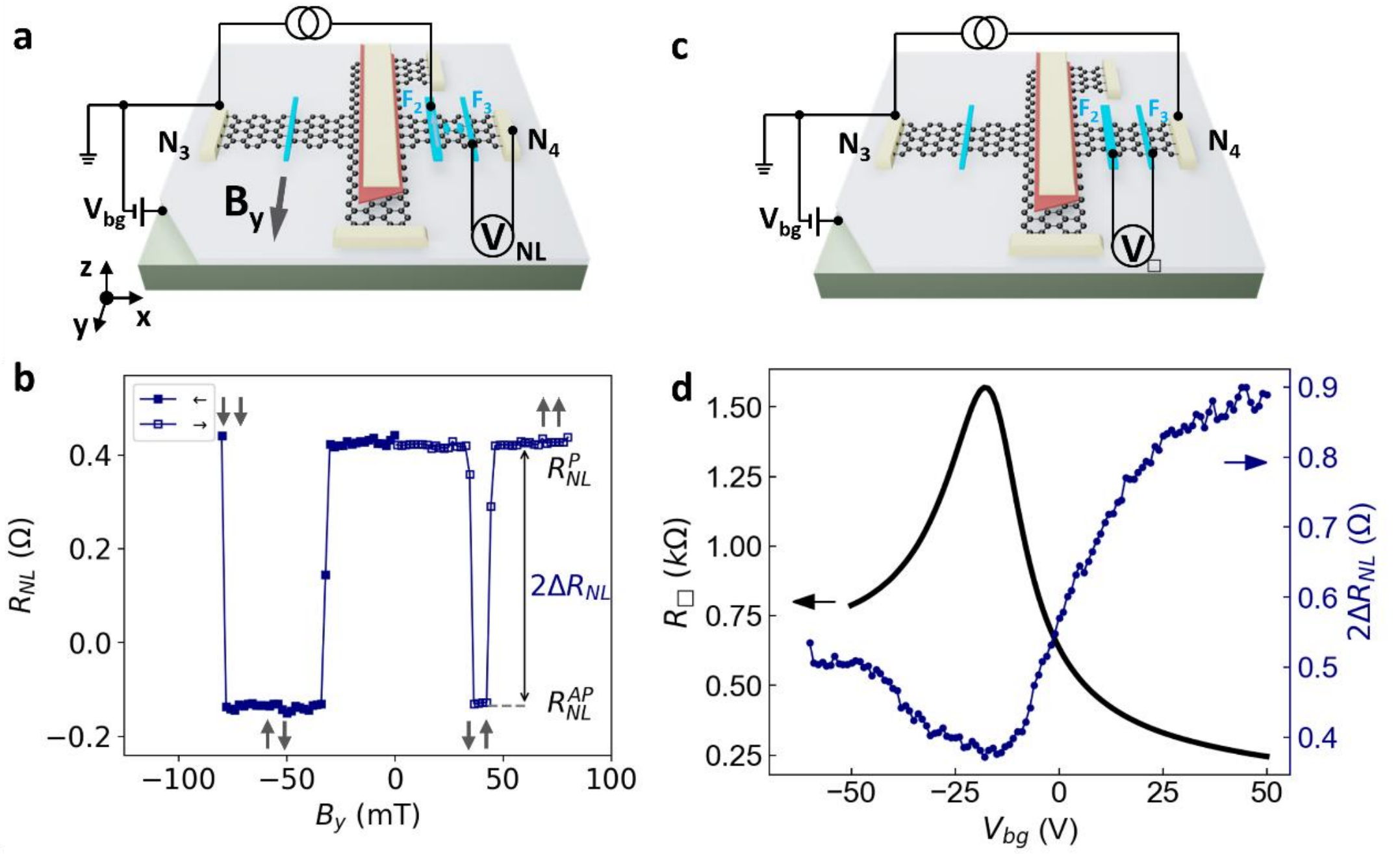

**Fig. 2 Charge and spin transport in the pristine graphene. a,** Sketch of the nonlocal spin valve measurement geometry used to characterize spin transport in the pristine graphene region. A charge current, $I_c$, is injected between the ferromagnetic contact $F_2$ and the non-magnetic contact $N_3$, while the nonlocal voltage, $V_{NL}$, is detected between $F_3$ and $N_4$. The nonlocal resistance is defined as $R_{NL} = V_{NL}/I_c$. An in-plane magnetic field, $B_y$, is applied along the easy axis of the ferromagnetic electrodes to switch their relative magnetization configuration. **b,** Nonlocal spin valve signal measured at 100 K as a function of $B_y$ at $V_{bg} = 0\,V$, showing the transition between parallel and antiparallel magnetization states, as indicated by the grey arrows. Filled and open symbols denote opposite magnetic-field sweep directions. The spin signal is defined as half of the difference in non-local resistance between the parallel $R_{NL}^{P}$ and antiparallel $R_{NL}^{AP}$ magnetization states, $\Delta R_{NL} = (R_{NL}^{P} - R_{NL}^{AP})/2$, as indicated by the vertical arrow. **c,** Sketch of the four-terminal charge-transport measurement configuration in the pristine graphene region. A charge current is applied between $N_3$ and $N_4$, and the longitudinal voltage is measured between $F_2$ and $F_3$ while the carrier density is tuned by the back-gate voltage, $V_{bg}$. **d,** Back-gate dependence of the graphene sheet resistance, $R_{\square}$, and nonlocal spin signal, $\Delta R_{NL}$, measured at 100 K. The sheet resistance shows a clear charge-neutrality point near $V_{bg} = -24$ V.

The heterostructure was then patterned into a Hall-bar geometry and contacted with normal-metal Pd/Au electrodes, labeled $N_1$–$N_4$, and ferromagnetic $TiO_2$/Co/Au tunnel contacts, labeled $F_1$–$F_3$, forming a nonlocal lateral spin valve device, as illustrated in Fig. 1e. The detailed fabrication process is described in Methods and Supporting Note 1. To independently control the carrier density and vertical electric field, a metal top-gate electrode was deposited on the native $HfO_2$ layer, while the highly doped Si substrate served as the global back gate. An optical image of the final dual-gated nonlocal lateral spin valve device is shown in Fig. 1f.

Before investigating the $HfS_2$/graphene vdW heterostructure region, we first characterize the charge and spin transport properties of the pristine graphene. Fig. 2a shows the non-local spin valve measurement configuration, where a charge current $I_{\mathrm{c}}$ is applied between the ferromagnetic electrode $F_2$ and the non-magnetic electrode $N_3$, while the non-local voltage $V_{\mathrm{NL}}$ is detected between $F_3$ and $N_4$. The nonlocal resistance is defined as $R_{\mathrm{NL}} = V_{\mathrm{NL}}/I_{\mathrm{c}}$. By sweeping the in-plane magnetic field, $B_y$, along the easy axis of the ferromagnetic electrodes, the magnetizations of $F_2$ and $F_3$ switch between parallel and antiparallel configurations owing to their different coercive fields. As shown in Fig. 2b, a clear two-level non-local resistance at 100 K, is observed, corresponding to the parallel $\left(R_{\mathrm{NL}}^{\mathrm{P}}\right)$ and antiparallel $\left(R_{\mathrm{NL}}^{\mathrm{AP}}\right)$ magnetization states. We further performed the Hanle spin precession measurements shown in Supporting Note 4. The pure spin signal is defined as $\Delta R_{\mathrm{NL}} = (R_{\mathrm{NL}}^{\mathrm{P}} - R_{\mathrm{NL}}^{\mathrm{AP}})/2$. We then characterize the charge transport of the same pristine graphene region using the four-terminal configuration shown in Fig. 2c. The current is applied between $N_3$ and $N_4$ and the voltage is detected between $F_2$ and $F_3$. The back-gate dependence of the sheet resistance, $R_{\square}$, exhibits a clear charge-neutrality point near $V_{\mathrm{bg}} = -24$ V, as shown in Fig. 2d. The gate dependence of the spin signal was measured by first preparing the magnetizations of $F_2$ and $F_3$ in either the parallel or antiparallel state, releasing the magnetic field to zero, and then recording $R_{\mathrm{NL}}$, with the raw data shown in Supporting Note 4. The resulting $\Delta R_{\mathrm{NL}}$ shows a pronounced carrier-density dependence, reaching a minimum near the charge-neutrality point and increasing at larger positive or negative gate voltages. This behavior is consistent with conventional graphene spin valve devices[21,22,47], where spin transport is strongly modulated by carrier density, and confirms that the pristine graphene region provides a robust reference channel for subsequent spin-transport measurements in the $HfS_2$/graphene vdW heterostructure.

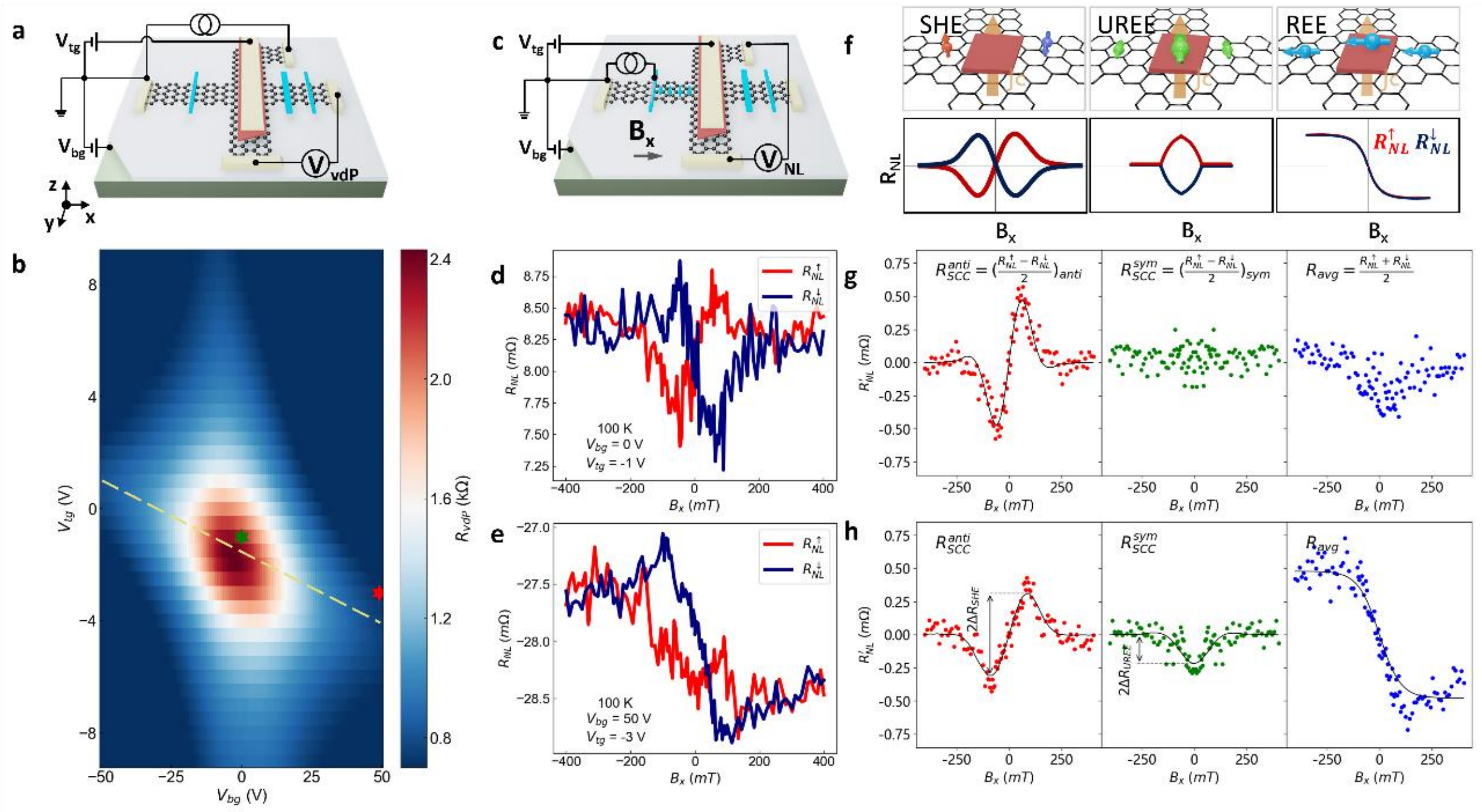

**Fig. 3 Electric-field control of spin–charge interconversion in $HfS_2$/graphene. a,** Measurement geometry for the van der Pauw-type resistance, $R_{\mathrm{vdP}}$, of the $HfS_2$-covered graphene region. **b,** $R_{\mathrm{vdP}}$ as a function of back-gate voltage, $V_{\mathrm{bg}}$, and top-gate voltage, $V_{\mathrm{tg}}$, measured at 100 K. The diagonal high-resistance feature, marked by the yellow dashed line, traces the charge-neutrality condition of the top-gated $HfS_2$/graphene region. The additional top-gate-insensitive response arises from uncovered graphene regions included in the current path. Two representative points near charge neutrality of the dual-gated region but with different vertical electric fields are selected for spin–charge interconversion measurements, marked with green and red stars, respectively. **c,** Non-local spin–charge conversion geometry. Spins are injected into graphene and diffuse into the $HfS_2$-covered region, where proximity-induced spin–orbit coupling converts the spin signal into a non-local voltage, $V_{\mathrm{NL}}$. An in-plane magnetic field, $B_x$, induces Hanle spin precession and rotates the injector magnetization. **d, e,** Measured non-local resistance traces, $R^{\uparrow}_{\mathrm{NL}}$ and $R^{\downarrow}_{\mathrm{NL}}$, obtained after initializing the injector magnetization along opposite $y$ directions for **d**, $V_{\mathrm{bg}} = 0$ V, $V_{\mathrm{tg}} = -1$ V, and **e**, $V_{\mathrm{bg}} = 50$ V, $V_{\mathrm{tg}} = -3$ V. **f,** Spin–charge interconversion mechanisms and the corresponding non-local resistance schematic resolved by the $B_x$-dependent symmetry. SHE converts out-of-plane spins, $s_z$, and may originate from Kane–Mele and/or valley-Zeeman SOC together with a staggered sublattice potential; the unconventional Rashba–Edelstein effect, UREE, converts the collinear in-plane spin component, $s_y$, associated with a radial Rashba field; and the conventional Rashba–Edelstein effect, REE, converts the transverse in-plane spin component, $s_x$, generated by the ordinary (tangential) Rashba field. **g, h,** Symmetry-resolved non-local resistance components $R'_{\mathrm{NL}}$ extracted from **d** and **e**, respectively. The REE and UREE responses are weak at small vertical electric field, whereas a pronounced REE signal emerges at larger vertical electric field, demonstrating gate control of Rashba spin–orbit coupling and spin–charge interconversion in $HfS_2$/graphene.

After confirming charge and spin transport in the pristine graphene region, we turn to the $HfS_2$-covered graphene channel, where spin–orbit coupling is induced by proximity. We first characterize the charge response of this proximitized region using the van der Pauw (vdP) configuration shown in Fig. 3a. Figure 3b shows the two-dimensional resistance map, $R_{\mathrm{vdP}}$, as a function of back-gate voltage, $V_{\mathrm{bg}}$, and top-gate voltage, $V_{\mathrm{tg}}$, measured at 100 K. A pronounced diagonal high-resistance feature is observed and is assigned to the charge-neutrality point (CNP) line of the $HfS_2$-covered graphene region, shown with the yellow dashed line in Fig. 3b. This diagonal dependence arises because the carrier density in the covered graphene is controlled simultaneously by the back gate and the local top gate, so that the two gate-induced charge densities compensate along the charge-neutrality line. Raman spectroscopy confirms the trilayer thickness of the graphene, while its stacking order was not specifically assigned. ABA- and ABC-stacked trilayer graphene can exhibit different displacement-field responses, with a field-induced gap expected primarily for ABC stacking[48]. No clear gap-opening signature is observed at 100 K, possibly because thermal activation and disorder broadening in the non-encapsulated, $SiO_2$-supported graphene, together with the uncovered-graphene contribution, weaken the associated resistance enhancement[49,50]. To separate carrier-density and vertical-electric-field effects, this dual-gate resistance map can be used as a reference: points along the diagonal charge-neutrality trajectory have comparable carrier-density conditions but different gate-bias asymmetry across the heterostructure. The CNP trajectory can be approximated as $V_{tg}^{CNP} = -0.06V_{bg} - 2$, giving $|\, dV_{tg}^{CNP}/dV_{bg} \,| = C_{bg}/C_{tg} = 0.06$, where $C_{bg}$ and $C_{tg}$ are the areal capacitances of the $SiO_2$ back-gate dielectric and the top $HfO_2$ dielectric, respectively. Taking $C_{bg} = 1.15 \times 10^{-4}\ \mathrm{F\,m^{-2}}$ for 300 nm $SiO_2$, we estimate $C_{tg} = 1.92 \times 10^{-3}\ \mathrm{F\,m^{-2}}$. The corresponding approximate vertical-field scale is then estimated as $E_z^{eff} = [C_{tg}(V_{tg} + 2) - C_{bg}V_{bg}]/(2\varepsilon_0)$. For the approximately 25 nm oxidized top layer, this corresponds to an effective dielectric constant of $\varepsilon_{r,top}^{eff} \approx 5.4$. This value is lower than typical values above 10 reported for dense $HfO_2$[42,51], which we attribute to native oxidation and residual sulfur in the oxidized region. The measurement geometry also probes uncovered graphene regions that are outside the top-gated $HfS_2$/graphene area. These regions are mainly controlled by $V_{\mathrm{bg}}$, giving rise to a top-gate-insensitive contribution in the resistance map.

We next probe spin–charge interconversion in the $HfS_2$-covered graphene region using the non-local spin precession geometry shown in Fig. 3c. To disentangle the role of the vertical electric field from carrier-density effects, we choose two representative gate configurations close to the CNP line: $V_{\mathrm{bg}} = 0\mathrm{V}$, $V_{\mathrm{tg}} = -1\mathrm{V}$, corresponding to a small effective vertical field of $E_z^{\mathrm{eff}} \approx +0.11\ \mathrm{V\ nm^{-1}}$, and $V_{\mathrm{bg}} = 50\mathrm{V}$, $V_{\mathrm{tg}} = -3\mathrm{V}$, corresponding to a larger vertical electric field $E_z^{eff} \approx -0.43\ \mathrm{V\,nm^{-1}}$. A charge current is injected from the ferromagnetic contact into the graphene channel, generating a spin accumulation that diffuses into the proximitized $HfS_2$/graphene region. The resulting non-local voltage, $V_{\mathrm{NL}}$, is detected across the $HfS_2$-covered Hall-bar region and normalized as $R_{\mathrm{NL}} = V_{\mathrm{NL}}/I_{\mathrm{c}}$. For each gate condition, the injector magnetization is first initialized along $+y$ or $-y$, giving $R_{\mathrm{NL}}^{\uparrow}$ and $R_{\mathrm{NL}}^{\downarrow}$, respectively. The magnetic field $B_x$ is then swept along the in-plane hard axis. This field induces Hanle precession of the injected $s_y$ spins in the $y$–$z$ plane and, at larger $B_x$, gradually rotates the injector magnetization towards the $x$ direction. Therefore, the measured $R_{\mathrm{NL}}^{\uparrow}$ and $R_{\mathrm{NL}}^{\downarrow}$ signals contain distinct contributions from spin components $s_z$, $s_x$ and $s_y$, corresponding to SHE, REE and UREE, respectively, which can be separated by their symmetry with respect to $B_x$ and to the injector magnetization. The SHE is commonly associated with valley-Zeeman SOC in graphene/TMDC vdW heterostructures[13], where opposite out-of-plane spin polarizations are locked to opposite valleys. It converts a spin current with $s_z$ polarization into a transverse charge voltage. A Kane–Mele contribution together with a staggered sublattice potential may also contribute to the spin–orbit field, while the present symmetry analysis identifies the $s_z$-polarized SCC component without directly separating these microscopic terms[6,13,38]. In our geometry, $s_z$ is generated from the injected $s_y$ spins by Hanle precession around $B_x$, resulting in an antisymmetric Hanle response that reverses sign with the injector magnetization. The REE arises from the conventional tangential Rashba field induced by out-of-plane inversion-symmetry breaking[4,5,37] at the $HfS_2$/graphene interface and converts the transverse in-plane spin component, $s_x$, into a charge signal. Since $s_x$ is mainly produced when $B_x$ rotates the injector magnetization towards the hard axis, this contribution shows a field-odd, S-shaped dependence. In contrast, the UREE originates from a reduced-symmetry or twist-induced radial Rashba field[8,18,19,33,38] and converts the collinear in-plane spin component, $s_y$, which is directly injected at zero field. This component is reduced by Hanle precession under $B_x$, giving an even-in-$B_x$ Hanle-like response.

The raw spin–charge interconversion traces measured at the two representative gate points are shown in Fig. 3d,e. Although both points are chosen close to the charge-neutrality condition of the $HfS_2$-covered graphene region, the $B_x$-dependent nonlocal signals already show markedly different symmetries. At $V_{\mathrm{bg}} = 0$ V and $V_{\mathrm{tg}} = -1$ V, where the vertical electric field is small, the two traces $R_{\mathrm{NL}}^{\uparrow}$ and $R_{\mathrm{NL}}^{\downarrow}$ mainly display an antisymmetric Hanle-like response. In contrast, at $V_{\mathrm{bg}} = 50$ V and $V_{\mathrm{tg}} = -3$ V, where the vertical electric field is much larger, the raw traces acquire a pronounced field-odd, S-shaped component, indicating a strong contribution from in-plane Rashba-type spin–charge interconversion.

The symmetry-resolved components of the non-local resistance $R'_{\mathrm{NL}}$ are plotted in Fig. 3g,h. The decomposition was performed directly from the two $B_x$-dependent traces measured after initializing the injector magnetizations in opposite directions, $R_{\mathrm{NL}}^{\uparrow}$ and $R_{\mathrm{NL}}^{\downarrow}$. Their average, $R_{ave} = (R_{\mathrm{NL}}^{\uparrow} + R_{NL}^{\downarrow})/2$, gives the injector-independent response, which contains the REE-related contribution together with residual spin-independent backgrounds[52]. Their difference isolates the spin-dependent SCC signal $R_{SCC} = (R_{\mathrm{NL}}^{\uparrow} - R_{NL}^{\downarrow})/2$. This difference signal is then separated into its $B_x$-antisymmetric part $R_{SCC}^{anti}$, associated with the SHE contribution, and its $B_x$-symmetric part $R_{SCC}^{sym}$, associated with the UREE contribution. At small vertical electric field, the signal is dominated by a clear SHE component, while both Rashba-related contributions are below the detection limit: neither a field-odd S-shaped REE signal nor a field-symmetric UREE signal can be resolved. The black solid curve in Fig. 3g shows the fit of the SHE component using the spin precession diffusion model, yielding an effective spin Hall angle of $\theta_{\mathrm{SHE}} = 0.012 \pm 0.006$, spin-diffusion coefficient $D_s = (2.8 \pm 0.3) \times 10^{-3} m^2 s^{-1}$ and the spin lifetime $\tau_s = 33.3 \pm 3.1\, ps$, corresponding to a spin diffusion length in the proximitized graphene of $\lambda_s^{\mathrm{prox}} = 306 \pm 20$ nm, with fitting details shown in Supporting Note 9. Consistently, additional SCC measurements in Device 2, performed along gate trajectories where either $V_{tg}$ or $V_{bg}$ is varied while the other gate is held at 0 V, show an SHE-dominated antisymmetric Hanle-like response without resolvable REE or UREE contributions (Supporting Note 7). Reciprocal charge-to-spin conversion measurements in Device 2 yield an antisymmetric Hanle response with the same field dependence and a comparable amplitude to the SCC response within experimental uncertainty, consistent with Onsager reciprocity[18] (Supporting Note 7). This indicates that, near zero vertical electric field, the in-plane Rashba spin texture is strongly suppressed and the SCC response is mainly governed by the out-of-plane polarized SHE channel. In contrast, at large vertical electric field, the SHE component remains clearly visible, but two additional Rashba-related contributions emerge. The REE exhibits a pronounced field-odd S-shaped dependence, consistent with the conversion of the transverse in-plane spin component $s_x$ generated by magnetization rotation under $B_x$. At the same time, the UREE becomes detectable as a field-symmetric SCC component with a finite zero-field amplitude, consistent with the direct conversion of the injected collinear spin polarization $s_y$. Fitting the symmetry-resolved components plotted with the solid black lines in Fig. 3h gives $\theta_{\mathrm{SHE}} = 0.005 \pm 0.002$, $D_s = (5.0 \pm 1.7) \times 10^{-3} m^2 s^{-1}$ and the spin lifetime $\tau_s = 23.0 \pm 8.0\, ps$, corresponding to $\lambda_s^{\mathrm{prox}} = 340 \pm 80$ nm for the SHE component, together with effective Rashba-related conversion efficiencies of $\alpha_{\mathrm{REE}} = 0.007 \pm 0.003$ and $\alpha_{\mathrm{UREE}} = 0.004 \pm 0.003$. The appearance of both REE and UREE only at large vertical electric field indicates that the vertical field activates or enhances the in-plane Rashba spin texture in $HfS_2$/graphene, enabling electrical control over both conventional and unconventional Rashba–Edelstein spin–charge interconversion. The increased carrier diffusivity at large vertical electric field partly compensates the shorter effective spin lifetime. The reduction in $\tau_s$, together with the emergence of the REE and UREE, is consistent with enhanced Rashba-induced spin relaxation at large vertical field. Additional gate-dependent SCC traces in the same device further show that carrier-density tuning does not qualitatively change the extracted SCC symmetry: in the low-vertical-field regime the response remains SHE-dominated, whereas in the high-vertical-field regime the Rashba-related REE and UREE components persist while their amplitudes are modulated (Supporting Note 5). The SOC underlying the SHE may contain both valley-Zeeman-like and Kane–Mele-like contributions, whereas the large-field REE and UREE directly reflect the enhanced in-plane Rashba spin texture.

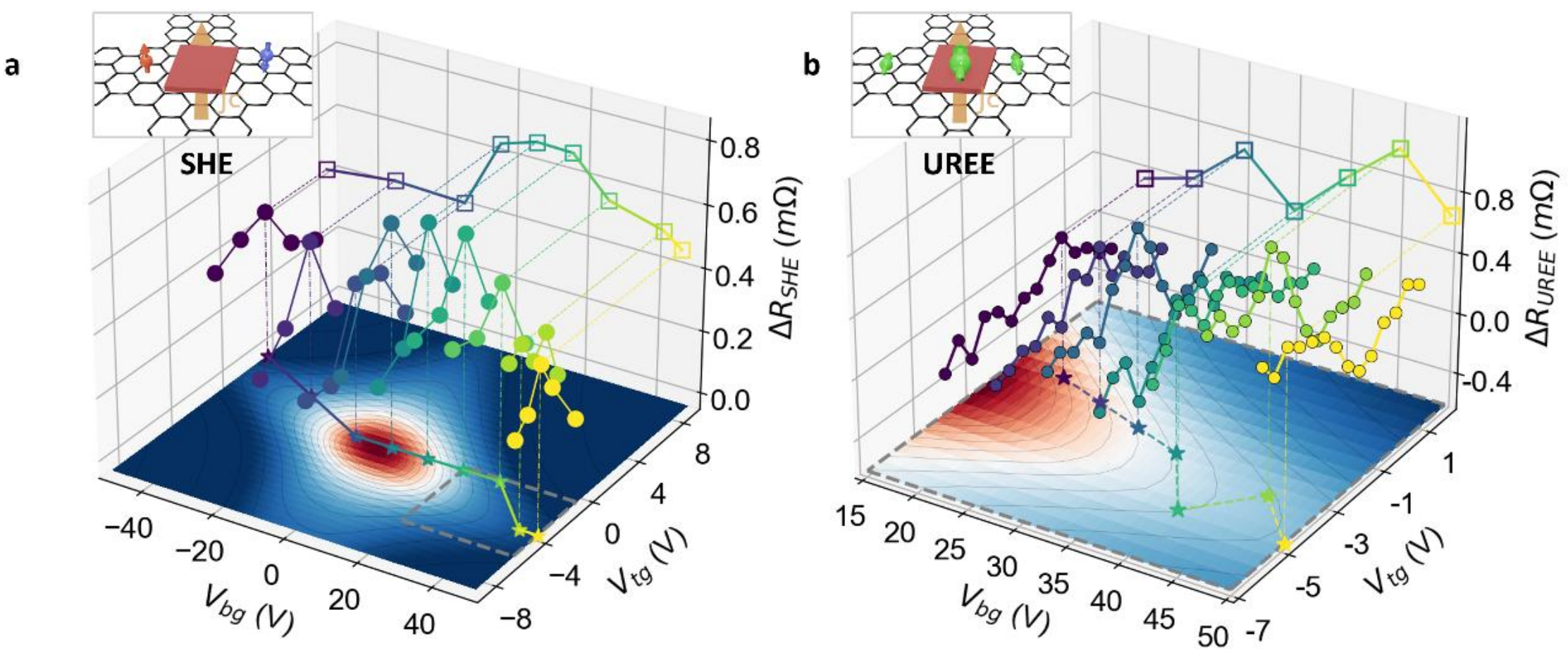

**Fig. 4 Dual-gate mapping of SHE and UREE in $HfS_2$/graphene. a,** Gate dependence of the extracted SHE amplitude, $\Delta R_{\mathrm{SHE}}$, obtained from $B_x$-dependent SCC measurements. The color map projected onto the $V_{\mathrm{bg}}$–$V_{\mathrm{tg}}$ plane shows the raw $R_{\mathrm{vdP}}$ of the $HfS_2$-covered graphene region, providing the charge-transport reference for the CNP trajectory. The maximum SHE response follows this trajectory, indicating enhanced out-of-plane polarized SCC near charge neutrality. The gray dashed rectangle marks the gate window selected for the detailed UREE mapping in **b**. **b,** Gate dependence of the UREE amplitude, $\Delta R_{\mathrm{UREE}}$, extracted from $B_y$-dependent magnetization-switching measurements. The bottom projection shows the raw $R_{\mathrm{vdP}}$ map in the selected gate window. The UREE is enhanced near the CNP trajectory but becomes pronounced only at larger gate biases, where the vertical electric field is stronger, consistent with electric-field activation of the in-plane Rashba spin texture.

Having identified the field-dependent SCC symmetry at selected gate configurations, we next map the corresponding SCC amplitudes across the dual-gate plane. Figure 4a shows the gate dependence of the SHE amplitude, $\Delta R_{\mathrm{SHE}}$, obtained from the $B_x$-dependent symmetry analysis. The color map projected onto the $V_{\mathrm{bg}}$–$V_{\mathrm{tg}}$ plane shows the raw $R_{\mathrm{vdP}}$ map of the $HfS_2$-covered graphene region, providing the charge-transport reference for the CNP trajectory. The extracted $\Delta R_{\mathrm{SHE}}$ follows this trajectory and reaches its maximum value of $0.71\ \mathrm{m\Omega}$ at $V_{\mathrm{bg}} = 10\ \mathrm{V}$ and $V_{\mathrm{tg}} = -1\ \mathrm{V}$. At the high-gate-bias point $V_{\mathrm{bg}} = 50\ \mathrm{V}$and $V_{\mathrm{tg}} = -3\ \mathrm{V}$, $\Delta R_{\mathrm{SHE}}$ decreases to $0.46\ \mathrm{m\Omega}$. An additional stacked two-dimensional line-cut representation of $\Delta R_{\mathrm{SHE}}$, plotted together with the normalized $R_{\mathrm{vdP}}$ CNP reference at each $V_{\mathrm{bg}}$, is provided in Supporting Note 5 to display the measured SCC amplitudes and their gate dependence. The extracted $\Delta R_{\mathrm{SHE}}$, which can also be affected by gate-dependent transport parameters, follows the CNP trajectory, suggesting that the SHE-like response is favored near charge neutrality and is consistent with an enhancement of out-of-plane spin Hall conversion[13,38].

We then examine the gate dependence of the Rashba-related SCC channels. For a more detailed gate mapping, we use the UREE as a representative Rashba-related signal using the $B_y$-dependent measurement, with the Co injector magnetization switching between opposite $y$ directions and reversing the injected $s_y$ polarization. The corresponding step-like SCC response gives $\Delta R_{\mathrm{UREE}}$, with representative raw data shown in Supporting Note 6. The resulting $\Delta R_{\mathrm{UREE}}$ map is shown in Fig. 4b. The gray dashed rectangle in Fig. 4a marks the gate window selected for this UREE mapping. Outside this window, the UREE response is below the experimental resolution. The bottom projection shows the raw $R_{\mathrm{vdP}}$ map in the selected gate window, providing the charge-transport reference for the CNP trajectory. The UREE response is enhanced around this trajectory, indicating that efficient Rashba-related SCC is favored when the Fermi level is tuned near charge neutrality. Within this window, $\Delta R_{\mathrm{UREE}}$ ranges from $0.34\ \mathrm{m\Omega}$ at $V_{\mathrm{bg}} = 20\ \mathrm{V}$, $V_{\mathrm{tg}} = -1.375\ \mathrm{V}$, to $1.00\ \mathrm{m\Omega}$ at $V_{\mathrm{bg}} = 45\ \mathrm{V}$, $V_{\mathrm{tg}} = -3.375\ \mathrm{V}$. The UREE amplitude is enhanced around the CNP trajectory and increases toward larger positive $V_{\mathrm{bg}}$ and more negative $V_{\mathrm{tg}}$, where the vertical-field asymmetry is stronger. The corresponding raw maps are shown in Supporting Note 6. These results indicate that efficient Rashba-related SCC requires both carrier-density tuning near charge neutrality and a sufficiently strong vertical electric field to activate the in-plane Rashba spin texture.

The opposite gate evolution of the SHE and Rashba-related SCC components provides further evidence that the dual gate changes the spin texture, rather than merely rescaling the spin signal. The measurements use the same non-local injection geometry, so a purely transport-related change would mainly rescale the SCC amplitude. Instead, the SHE is maximized near the CNP trajectory and decreases in the large-vertical-field regime, whereas the REE and UREE become detectable and increase under large vertical field. This behavior is consistent with theoretical work on graphene/TMDC vdW heterostructures, where the out-of-plane and in-plane Rashba-type SOC terms respond differently to gating, with transverse electric fields primarily enhancing the Rashba contribution[19,37,38]. A dual-gated hBN/graphene control device exhibits clear gate modulation and spin transport but no reproducible SCC under representative gate conditions, supporting the $HfS_2$-proximity origin of the SCC observed here (Supporting Note 8). The data therefore indicate that increasing the vertical electric field drives the proximitized graphene from a mainly out-of-plane spin texture towards a stronger in-plane Rashba spin texture.

## CONCLUSIONS

To conclude, our results show that dual gating of $HfS_2$/graphene modifies the relative contribution of different spin-charge interconversion channels, rather than simply changing the overall signal amplitude. This conclusion is based on three experimental observations. First, the dual-gate charge-transport map identifies the charge-neutrality trajectory of the $HfS_2$-covered graphene, providing a reference for separating carrier-density effects from vertical-electric-field effects. Second, Hanle-precession-based symmetry analysis at representative gate configurations resolves the SCC response into components associated with $s_z$, $s_x$, and $s_y$ spin polarizations, corresponding to the SHE, REE and UREE channels, respectively. Third, the gate-dependent SCC maps show that the SHE amplitude is maximized near the charge-neutrality trajectory, whereas the Rashba-related response becomes more pronounced under stronger vertical electric fields. These results indicate that the vertical electric field changes the relative weight of out-of-plane and in-plane spin texture in proximitized graphene. Since spin-charge interconversion with different spin-polarization directions is becoming increasingly relevant for spin-orbit devices, the ability to electrically tune the balance between out-of-plane and in-plane spin-texture components provides a useful route for controlling SCC symmetry in vdW heterostructures.

## EXPERIMENTS AND METHODS

### Sample fabrication

Graphene flakes were mechanically exfoliated from natural graphite onto 300-nm-thick $SiO_2$ on doped Si substrates using Nitto tape and identified by optical contrast. Candidate flakes were characterized by micro-Raman spectroscopy to confirm the graphene thickness and crystalline quality, as shown in Supporting Note 2.

Bulk $HfS_2$ crystals (HQ Graphene) were mechanically exfoliated and transferred onto selected graphene flakes by a dry-transfer method. After locating suitable $HfS_2$ flakes under an optical microscope, the flakes were stamped onto graphene using a viscoelastic transfer tool with micrometer-controlled alignment. After the transfer, the upper part of $HfS_2$ was naturally oxidized under ambient conditions for 30 min, forming an $HfO_2$/$HfS_2$/graphene heterostructure. The oxidized $HfO_2$ layer serves as the top-gate dielectric, while the remaining $HfS_2$ layer provides the proximity interface to graphene.

After stamping, the $HfS_2$/graphene vdW heterostructure was nanopatterned into a Hall-bar geometry by electron-beam lithography using PMMA resist. A 20-nm-thick Al layer was deposited by thermal evaporation as a hard mask, followed by lift-off. The exposed regions were then removed by reactive-ion etching to define the active device channel. The Al hard mask was chemically removed using an MF developer containing tetramethylammonium hydroxide. Non-magnetic electrodes were fabricated by electron-beam lithography, deposition of 5 nm Pd and 45 nm Au by electron-beam evaporation, and lift-off in acetone.

The magnetic electrodes were fabricated by a final electron-beam lithography step. A 3 Å Ti layer was deposited and oxidized in air to form a $TiO_x$ tunnel barrier, followed by deposition of 35 nm Co and 15 nm Au capping layers. The $TiO_x$/Co electrodes were designed with different widths to obtain different coercive fields, allowing parallel and antiparallel magnetization configurations during spin valve measurements.

### Electrical measurements

Electrical measurements were performed in a Quantum Design Physical Property Measurement System using a DC reversal technique with a Keithley 2182 nanovoltmeter and a Keithley 6221 current source. The doped Si substrate was used as the global back-gate electrode, while the naturally oxidized $HfO_2$ layer was used as the top-gate dielectric. Gate voltages were applied using a Keithley 2636B source meter, with the leakage current less than 2 nA during all measurements. Non-local spin transport and spin-charge interconversion measurements were performed by injecting spins through $TiO_x$/Co ferromagnetic electrodes with a charge current of 80 $\mu A$ and detecting the voltage response in spatially separated non-magnetic contacts.

### Raman spectroscopy

Micro-Raman spectroscopy was performed at room temperature using a Renishaw inVia Qontor Raman system equipped with a 100× objective and a 2400 l/mm grating. A 532 nm laser was used for graphene characterization. The laser power was kept below 1 mW to avoid laser-induced damage. In the case of the graphene, the 2D peak was fitted with six Lorentzian components, confirming the trilayer nature of the graphene channel, while the weak D peak indicates low disorder after transfer and device fabrication[53].

**Cross-sectional STEM and EDS characterization**

Cross-sectional specimens were prepared from completed devices by focused-ion-beam milling. HAADF-STEM imaging and EDS elemental mapping were used to examine the vertical structure and elemental distribution of the $HfS_2$/graphene vdW heterostructure. The Hf and O maps confirm the formation of an oxidized Hf-based upper layer, while the residual S signal in the oxidized region indicates partial oxidation of $HfS_2$ rather than complete sulfur removal. The Pd and Au maps identify the top metal contact region.

## ASSOCIATED CONTENT

**Data and availability statement**

The data and analysis code supporting this study are available from the corresponding authors upon reasonable request.

**Supporting information**

The Supporting Information is available at the end of this manuscript.

## AUTHOR INFORMATION

**Author contributions**

H.Y. and F.C. conceived the study. H.Y. fabricated the samples, performed the electrical measurements and analyzed the data, with the help of E.D., Z.C., J.I.-A., and W.S. B.M.-G and H.Y. performed the Raman spectroscopy measurements. A.C. performed the STEM and EDS characterization. All authors contributed to the discussion of the results and their interpretation. H.Y. and F.C. wrote the manuscript with input from all authors.

**Competing interests**

The authors declare no competing interests.

## ACKNOWLEDGEMENTS

We acknowledge funding from MICIU/AEI/10.13039/474 501100011033 (Grants No. CEX2020-001038-M and CEX2025-001634-M), from MICIU/AEI and ERDF/EU (Project Nos. PID2024-155708OB-I00 and PID2021-128004NB-C21), and from the European Union's Horizon 2020 research and innovation program under the Marie Skłodowska-Curie Grant Agreement No. 955671. This work is also supported by MICIU and by the European Union NextGenerationEU/PRTR-C17.I1, as well as by IKUR Strategy under the collaboration agreement between Donostia International Physics Center and CIC nanoGUNE on behalf of the Department of Education of the Basque Government. H.Y. acknowledges support from the National Natural Science Foundation of China (Grant Nos. 62404014 and 62674010), Beijing Natural Science Foundation (Grant No. 4262054), and the Fundamental Research Funds for the Central Universities (501XYGG2026142003). B.M.-G. and M.G. thank support from "Ramon y Cajal" Programme by the Spanish MICIU/AEI/10.13039/501100011033 and European Union NextGenerationEU/PRTR (Grants No. RYC2021-034836-I and RYC2021-031705-I, respectively).

## SUPPORTING INFORMATION for

# Electric-field tunable spin texture in dual-gated $HfS_2$/graphene van der Waals heterostructures

**Note 1. Optical micrographs of the device fabrication sequence.**

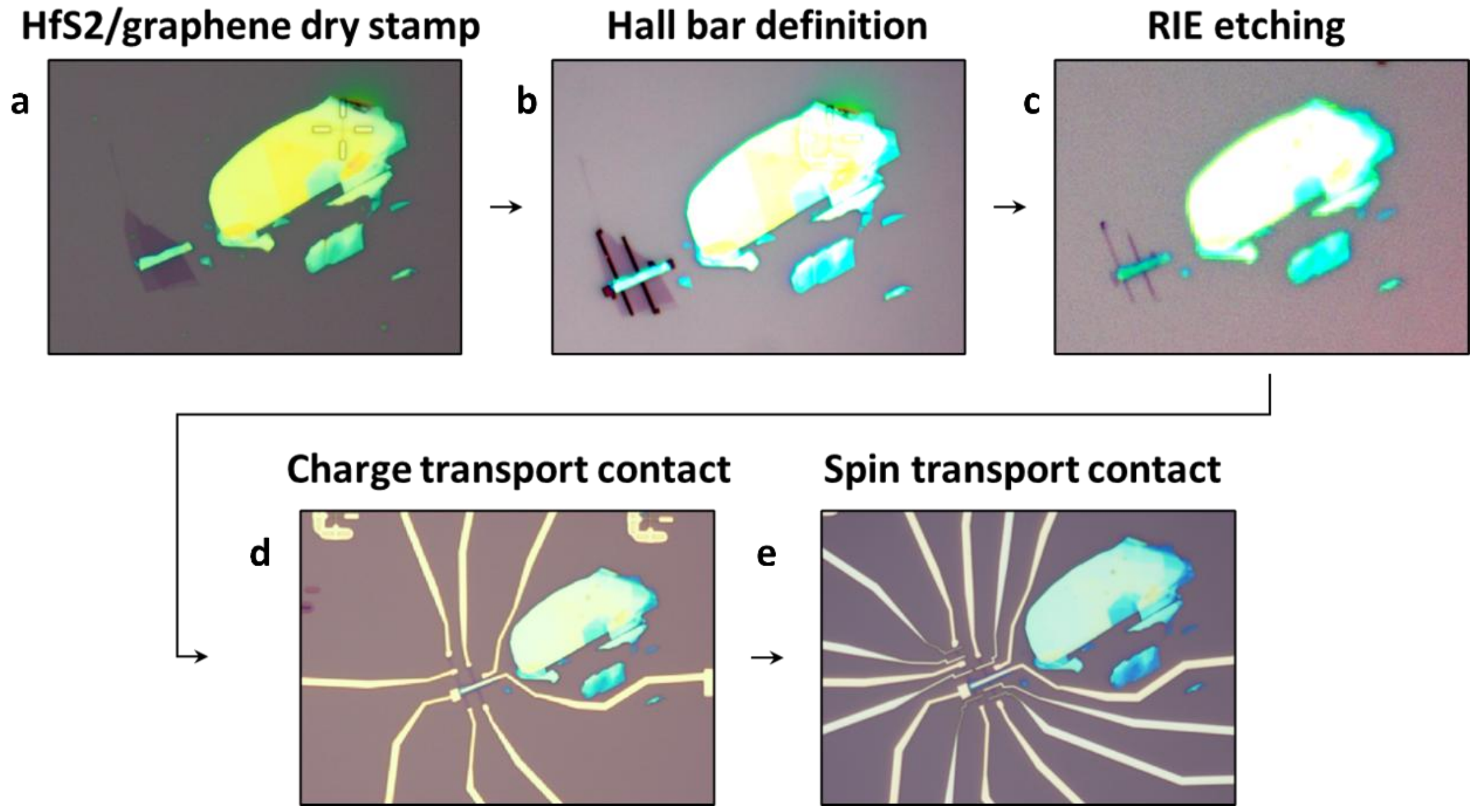

**Fig. S1 Fabrication process of the $HfS_2$/graphene heterostructure device.** **a**, Optical micrograph of the $HfS_2$/graphene heterostructure after dry-transfer assembly on the $SiO_2$/Si substrate. **b**, Optical image after defining the Hall-bar mesa region using PMMA as the etching mask, following electron-beam lithography and development. **c**, Device after deposition and lift-off of an Al hard mask, reactive-ion etching of the exposed heterostructure region, and subsequent removal of the Al mask using MF developer. **d**, Optical image after fabrication of the non-magnetic Pd/Au electrodes for charge and nonlocal spin-transport measurements. **e**, Final device after deposition of $TiO_2$ tunnel barriers and Co ferromagnetic contacts, completing the lateral spin valve geometry used for spin injection, detection and spin-charge conversion measurements.

**Note 2. Raman characterization of the trilayer graphene spin channel.**

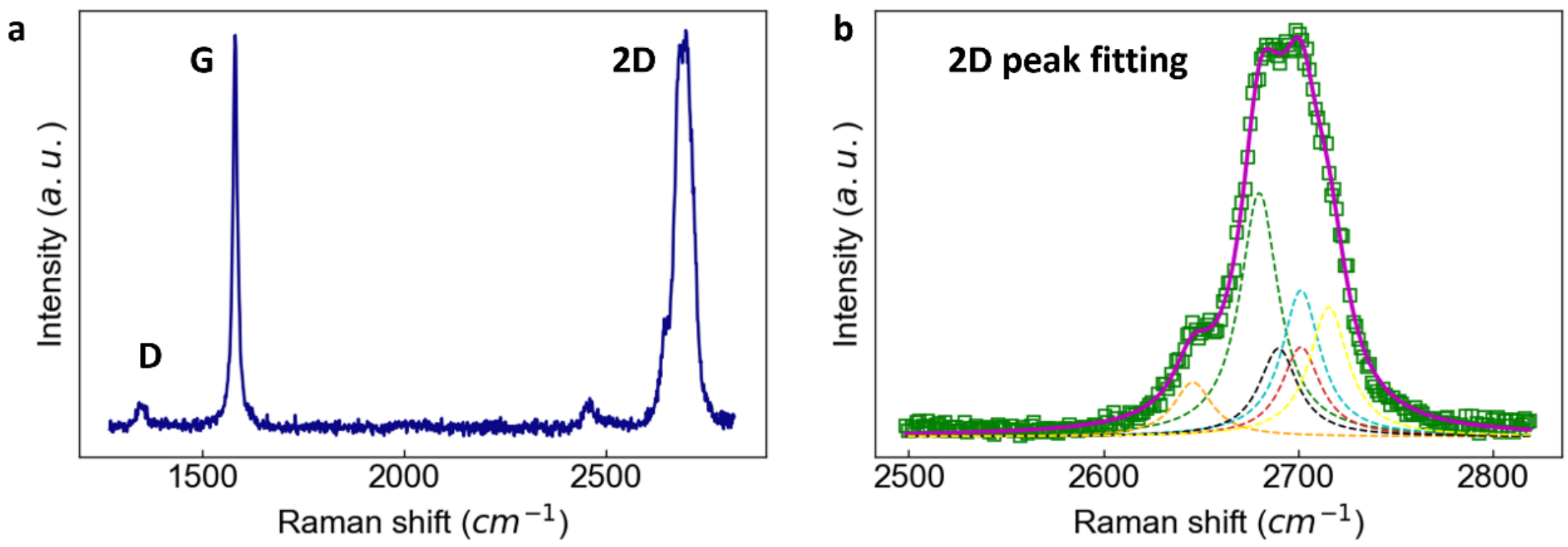

**Fig. S2 Raman characterization of the trilayer graphene channel. a**, Representative Raman spectrum measured from the graphene region of the $HfS_2$/graphene heterostructure, showing the characteristic D, G and 2D bands. The pronounced G and 2D peaks identify the graphene channel, while the relatively weak D peak indicates a low defect density and limited damage introduced during transfer and device fabrication. **b**, Enlarged 2D-band spectrum fitted with six Lorentzian components. The multi-component 2D-band line shape is consistent with trilayer graphene, confirming the layer number of the spin-transport channel used in the device.

**Note 3. EDS elemental mapping of the cross-sectional $HfS_2$/graphene heterostructure.**

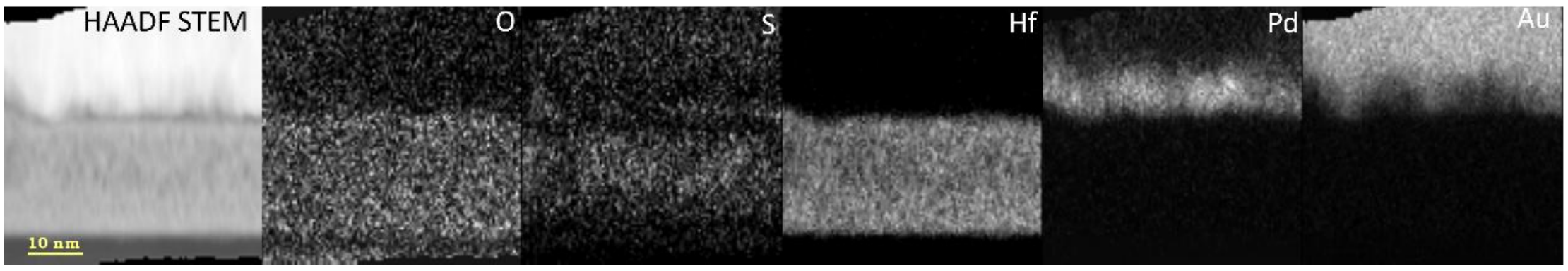

**Fig. S3 Cross-sectional scanning transmission electron microscopy (STEM) and energy dispersive spectroscopy (EDS) elemental mapping of the $HfS_2$/graphene heterostructure.** Cross-sectional HAADF-STEM (High-angle annular dark-field scanning transmission electron microscopy) image and corresponding EDS elemental maps of O, S, Hf, Pd, and Au for the $HfS_2$/graphene device. The elemental distributions reveal a layered structure across the heterostructure and top-contact region. The Hf signal clearly outlines the Hf-based layer, while the strong O signal indicates that the upper $HfS_2$ region is oxidized, consistent with the formation of $HfO_2$. The $HfS_2$/graphene heterostructure was exposed to ambient conditions for a fixed time of 30 min, while $HfS_2$ oxidation can progressively slow and exhibit time- and temperature-dependent self-limiting behavior[1]. The Pd and Au maps identify the metal top contact. We also note that S is still detected in the oxidized $HfO_2$ region, suggesting incomplete sulfur removal with the presence of residual sulfur in the partially oxidized $HfS_2$ layer, and the possible diffusing process when making the TEM sample cut.

**Note 4. Hanle spin precession analysis and gate-dependent spin signal in the pristine graphene channel.**

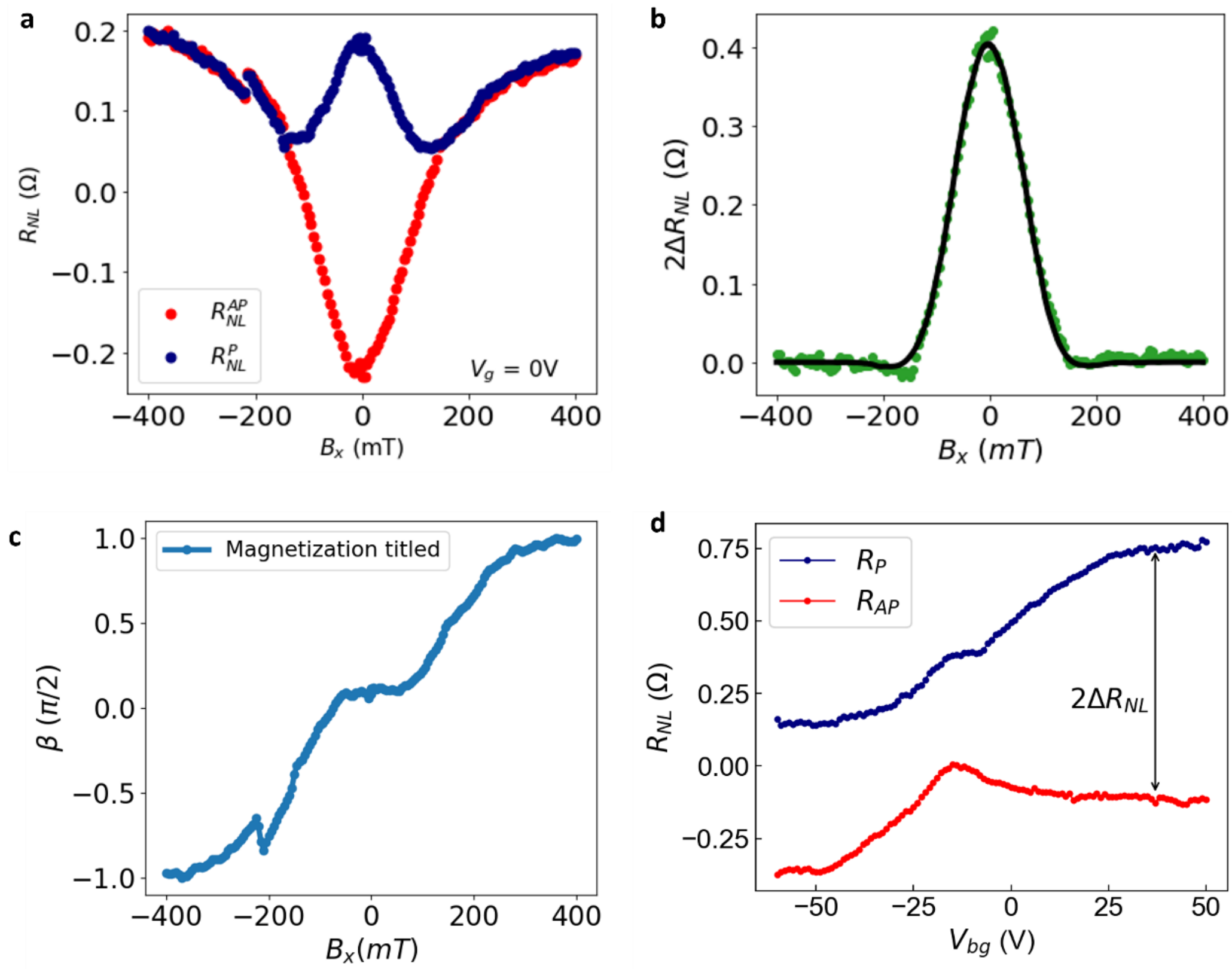

**Fig. S4 Spin-transport characterization of the pristine graphene reference channel. a,** Nonlocal Hanle spin precession traces measured at 300 K in the pristine graphene region at $V_{bg} = 0$ V. The blue and red curves correspond to the parallel and antiparallel magnetization configurations of the injector and detector electrodes, respectively. Sweeping the in-plane magnetic field $B_x$modulates the nonlocal resistance $R_{NL}$ through spin precession and dephasing during diffusion. **b,** Spin precession signal $2\Delta R_{NL}$, obtained from the difference between the parallel and antiparallel Hanle traces in **a**. The black curve shows the fit using the one-dimensional Bloch spin-diffusion model, taking into account the measured contact resistances of the $F_2$ and $F_3$ electrodes, $R_{c,F2} = 1.7$ kΩ and $R_{c,F3} = 3.1$ kΩ , determined from three-terminal measurements. The fitting gives a spin polarization $P = 0.07$, spin diffusion length $\lambda_s = 1.01$ μm, spin diffusion coefficient $D_s = 0.019$ m$^2$ s$^{-1}$, and spin lifetime $\tau_s = 5.4 \times 10^{-11}$ s. The pristine reference channel consists of graphene directly supported on $SiO_2$ and is not encapsulated. The extracted spin-diffusion length is of the same micrometer scale as that obtained in our previous non-encapsulated graphene spin-valve devices[2–4]. Its moderate value may arise from disorder in the exposed $SiO_2$-supported graphene, including charge inhomogeneity, substrate roughness, fabrication residues, and contact-induced spin relaxation[5]. **c,** The contact magnetizations progressively rotate from $y$ toward $x$ with $B_x$ applied along the hard axis. The corresponding field-dependent rotation factor[2], $\cos(\beta(B_x))$, was extracted from this measurement. The rotation approaches saturation above approximately 200 mT, consistent with the saturation field of the conventional REE response and the simultaneous suppression of the SHE and UREE components. **d,** Back-gate dependence of the raw nonlocal resistance in the parallel and antiparallel states at 100 K, $R_{NL}^{P}$and $R_{NL}^{AP}$, measured after initializing the two ferromagnetic electrodes into the corresponding magnetization configurations. The spin signal is extracted as $2\Delta R_{NL} = R_{NL}^{P} - R_{NL}^{AP}$, as indicated by the arrow. These raw data provide the basis for the gate-dependent spin signal shown in Fig. 2d of the main text.

**Note 5. Additional gate-dependent SCC traces and spin-texture decomposition.**

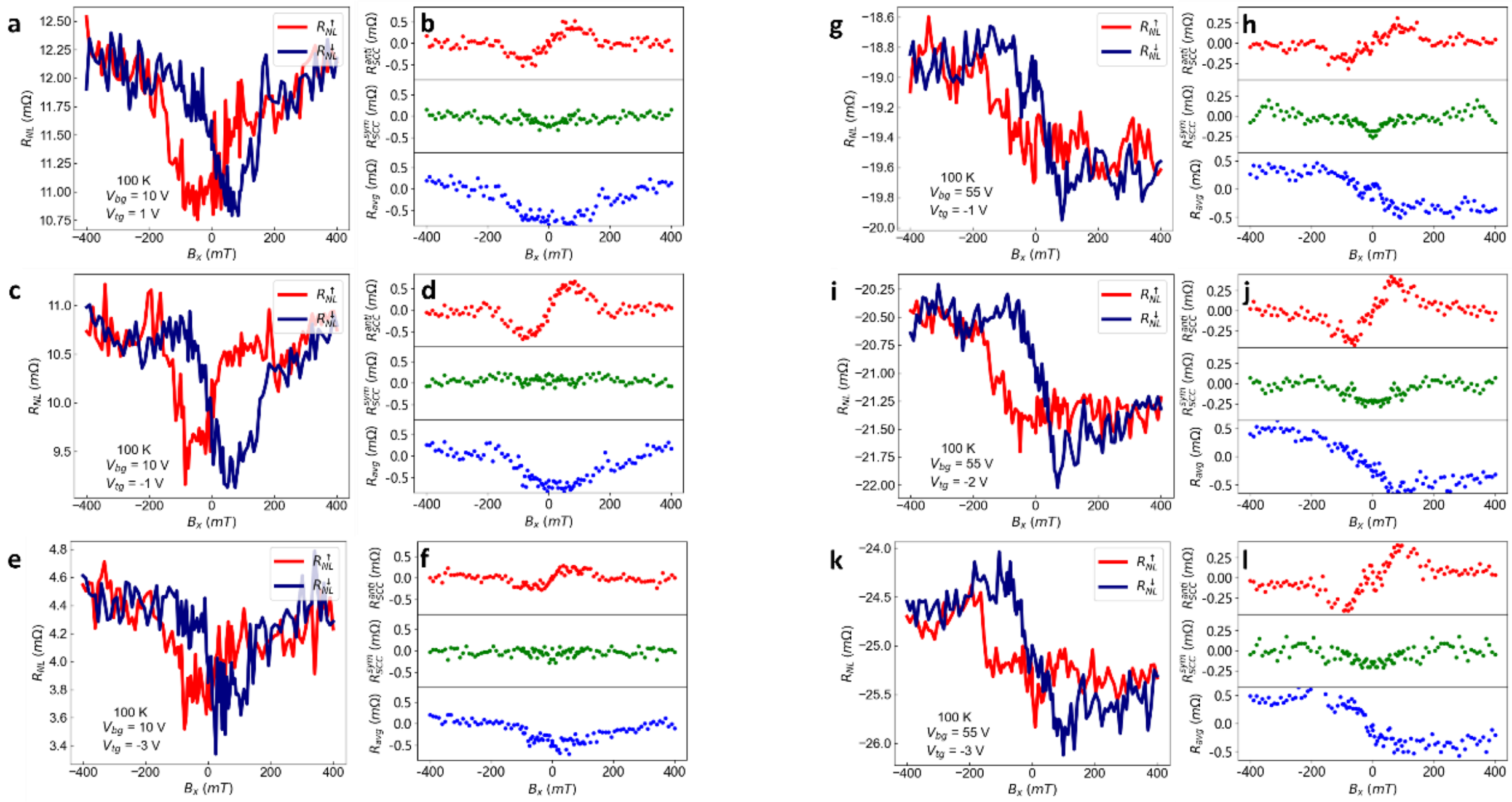

**Fig. S5 Raw SCC traces and extracted spin-charge conversion components under selected gate configurations. a,c,e,g,i,k**, Raw magnetic-field-dependent spin-charge conversion signals measured at 100 K under selected dual-gate conditions. The red and blue curves correspond to opposite spin-injection configurations measured as a function of the in-plane magnetic field $B_x$. **b,d,f,h,j,l**, Corresponding decomposition of the SCC response into three symmetry components associated with the Rashba-Edelstein effect, unconventional Rashba-Edelstein effect, and spin Hall effect. For the low-vertical-electric-field regime shown in **a-f** $\left(V_{\mathrm{bg}} = 10\ \mathrm{V}\right)$, changing $V_{\mathrm{tg}}$modifies the carrier density but does not qualitatively change the extracted SCC symmetry, indicating that carrier-density variation alone does not substantially alter the spin texture. In contrast, in the high-vertical-electric-field regime shown in **g-l** $\left(V_{\mathrm{bg}} = 55\ \mathrm{V}\right)$, a pronounced Rashba-like SCC response is observed throughout the selected gate range. The SCC components are separated algebraically according to their transformation under injector-magnetization reversal and magnetic-field reversal. For each gate condition, the half-difference $\left[R_{NL}^{\uparrow}(B_x) - R_{NL}^{\downarrow}(B_x)\right]/2$ suppresses magnetization-independent offsets and common-mode drift. Its field-antisymmetric and field-symmetric parts correspond to the SHE and UREE components, respectively, while the field-antisymmetric part of the averaged trace $\left[R_{NL}^{\uparrow}(B_x) + R_{NL}^{\downarrow}(B_x)\right]/2$ gives the REE-related response. Random measurement noise does not systematically reproduce both the required magnetization-reversal and magnetic-field symmetries. These additional raw data support the conclusion that the emergence of the Rashba-like spin texture is primarily driven by the vertical electric field rather than by carrier-density modulation alone.

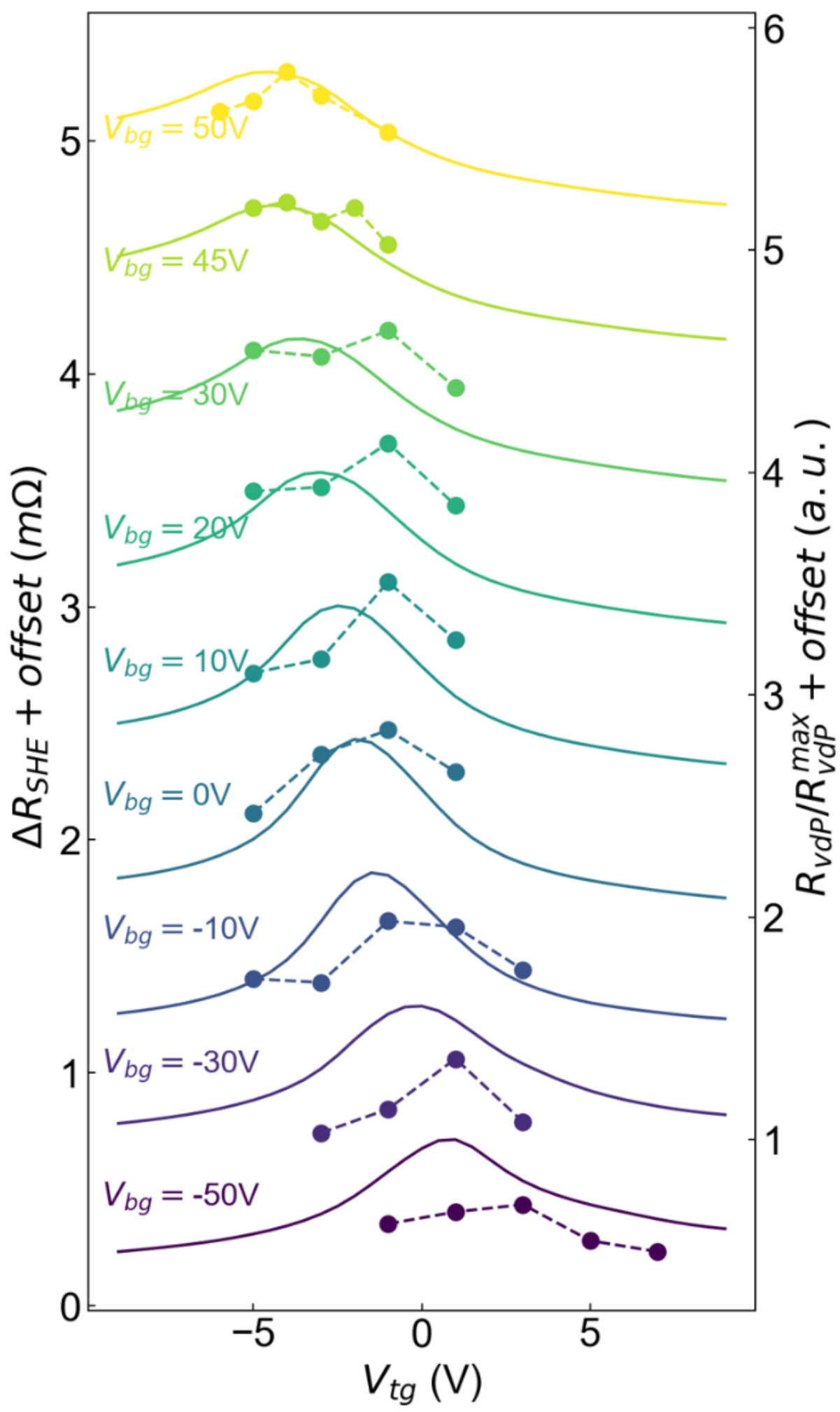

**Fig. S6 Stacked line-cut representation of the extracted SHE amplitude as a function of $V_{tg}$ at selected $V_{bg}$ values.** Filled circles connected by dashed lines show the measured $\Delta R_{SHE}$, with adjacent data sets vertically offset by 0.6 $m\Omega$ for clarity. Solid curves show the corresponding $R_{vdP}$ line cuts, each normalized to its own maximum and plotted on the right axis. The resistance maximum provides the CNP reference at each $V_{bg}$, allowing the gate evolution of the SHE response relative to the CNP trajectory to be directly compared.

**Note 6. $B_y$-scan extraction of the unconventional Rashba-Edelstein signal.**

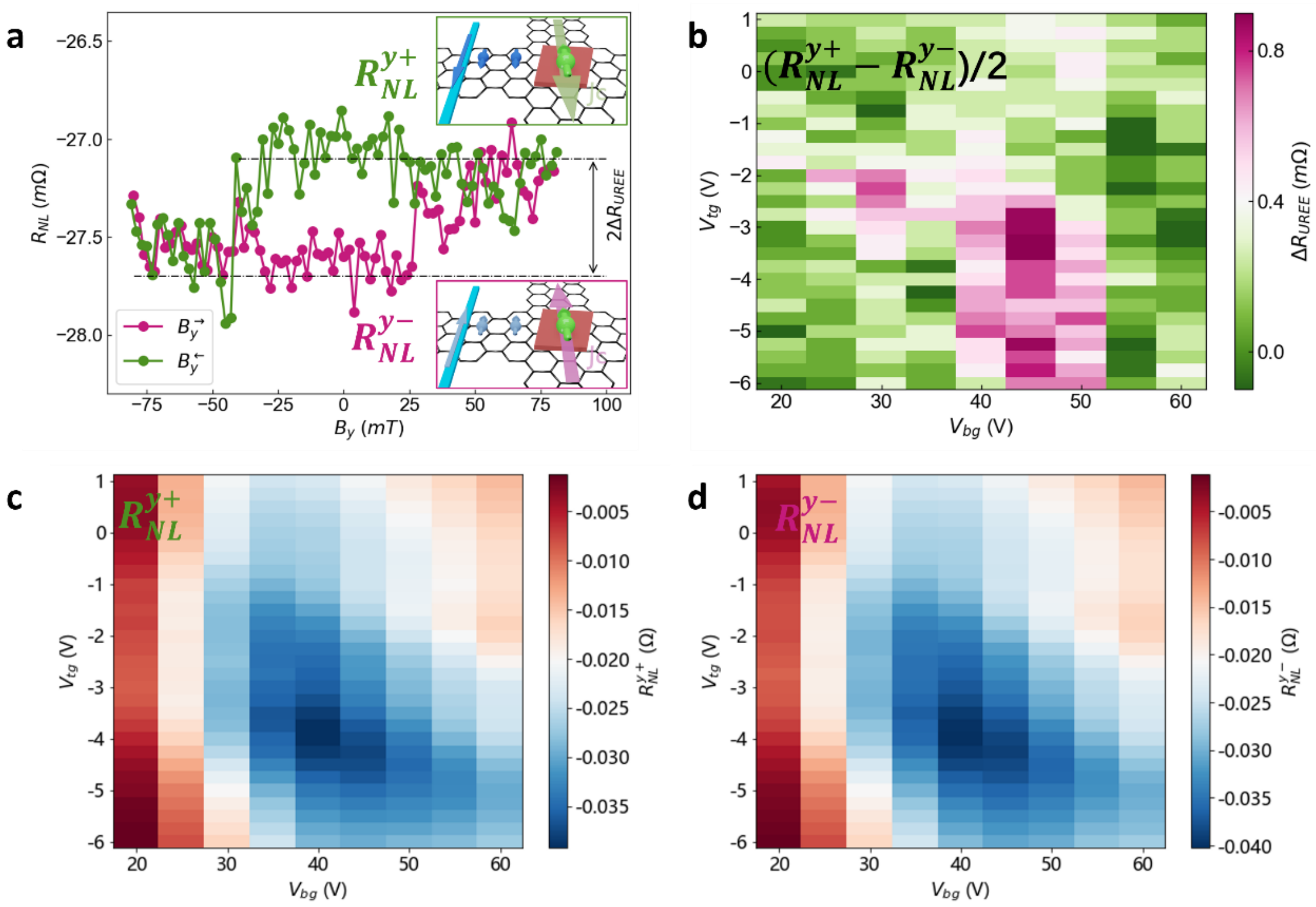

**Fig. S7 Representative $B_y$-scan trace and gate-dependent mapping of the UREE signal. a**, Representative $B_y$-scan of the nonlocal SCC resistance measured at a selected gate condition. Sweeping the in-plane magnetic field $B_y$ reverses the spin polarization injected into the $HfS_2$/graphene region, giving two non-local resistance levels, denoted as $R_{\mathrm{NL}}^{y+}$ and $R_{\mathrm{NL}}^{y-}$. The difference between these two levels defines the UREE-related SCC signal, $\Delta R_{\mathrm{UREE}} = (R_{\mathrm{NL}}^{y+} - R_{\mathrm{NL}}^{y-})/2$, as indicated in the figure. **c,d**, Gate-dependent raw maps of $R_{\mathrm{NL}}^{y+}$ and $R_{\mathrm{NL}}^{y-}$, respectively, extracted from $B_y$-scan measurements over the dual-gate parameter space. These two maps provide the direct experimental input before spin-dependent subtraction. **b**, Gate-dependent map of $\Delta R_{\mathrm{UREE}}$ obtained from the difference between the two raw SCC states. The subtraction suppresses spin-independent backgrounds and isolates the SCC component that changes sign with the injected spin polarization, corresponding to the unconventional Rashba-Edelstein contribution. The resulting map shows that the UREE signal is strongly gate dependent, consistent with an electric-field-tunable in-plane spin texture in the $HfS_2$/graphene heterostructure.

**Note 7. SCC measurements in Device 2.**

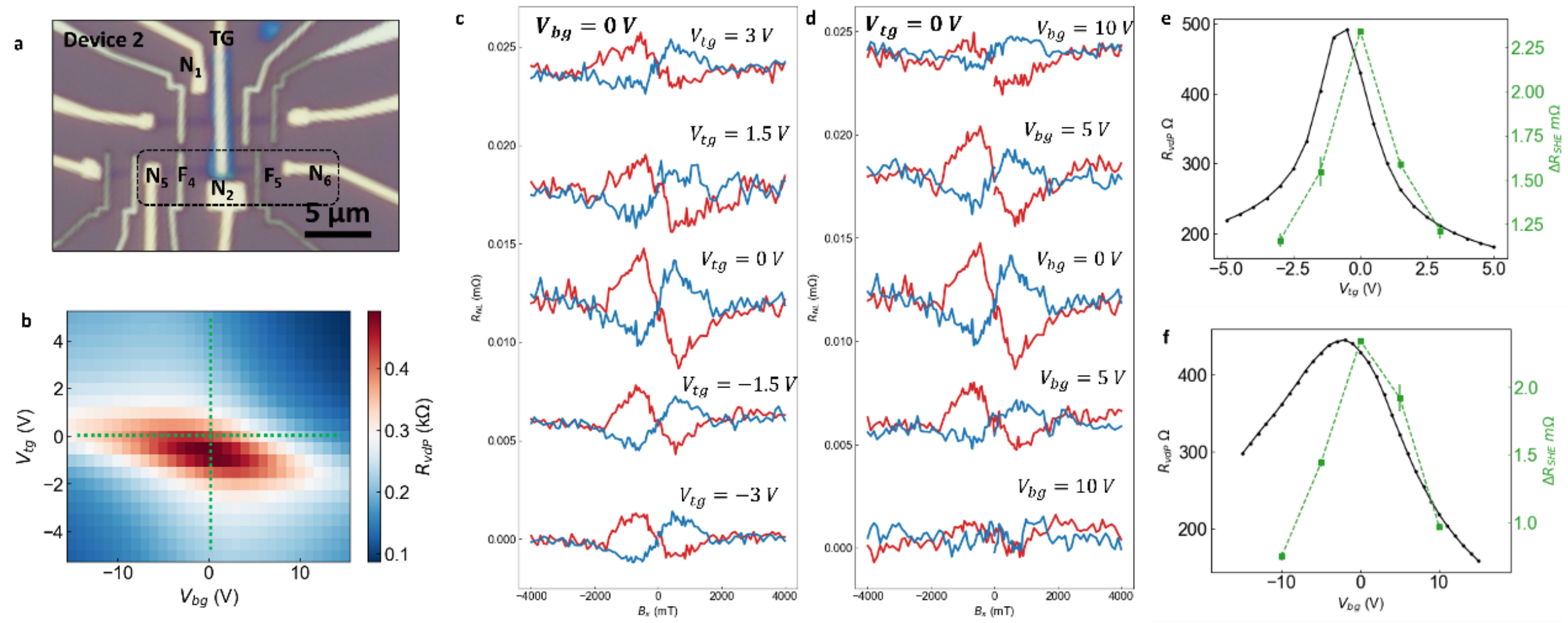

**Fig. S8. Gate-dependent spin–charge conversion response in Device 2. a,** Optical micrograph of Device 2, fabricated on another $HfS_2$/graphene region of the same sample. The graphene channel in this device is bilayer graphene. The top gate, non-magnetic electrodes and ferromagnetic spin-injection electrodes are labeled in the image. **b,** Dual-gate van der Pauw resistance map of the $HfS_2$-covered graphene region in Device 2, measured with current applied between $N_1$ and $N_5$ and voltage detected between $N_6$ and $N_2$. The diagonal high-resistance feature indicates the charge-neutrality trajectory of the top-gated $HfS_2$/graphene region, similar to the behavior observed in Device 1. Although a displacement-field-induced gap is generally expected in bilayer graphene, no clear systematic enhancement of the CNP resistance is resolved at 100 K. This may result from thermally activated transport and disorder-induced potential fluctuations or localized states in the $SiO_2$-supported, non-hBN-encapsulated graphene, together with the contribution of uncovered graphene regions to the measured resistance[6,7]. The dashed green lines mark the gate trajectories used for the SCC measurements in **c** and **d**. **c,** $B_x$-dependent SCC traces measured while varying $V_{tg}$ at $V_{bg} = 0$ V. The red and blue curves correspond to opposite spin-injection configurations and are vertically shifted for clarity. **d,** $B_x$-dependent SCC traces measured while varying $V_{bg}$ at $V_{tg} = 0$ V. Along both gate trajectories, the SCC response is dominated by an antisymmetric Hanle-like component, with no resolvable REE or UREE contribution under these conditions. **e,f,** Extracted SHE amplitude plotted together with the corresponding $R_{vdP}$ for the $V_{tg}$ and $V_{bg}$ sweeps, respectively. The SHE amplitude follows the gate-dependent charge response and is enhanced near the charge-neutrality region.

These measurements provide an additional device-level reference for the gate dependence of SCC in $HfS_2$/graphene. In Device 2, varying either $V_{tg}$ or $V_{bg}$ along the fixed-gate trajectories mainly modulates the amplitude of the SHE-dominated response, while the line shape remains governed by the antisymmetric Hanle-like component. The absence of a resolvable REE or UREE contribution under these conditions is consistent with the behavior in Device 1 away from the large vertical-field regime, supporting the interpretation that the emergence of Rashba-related in-plane SCC channels requires sufficient vertical-field asymmetry rather than carrier-density tuning alone.

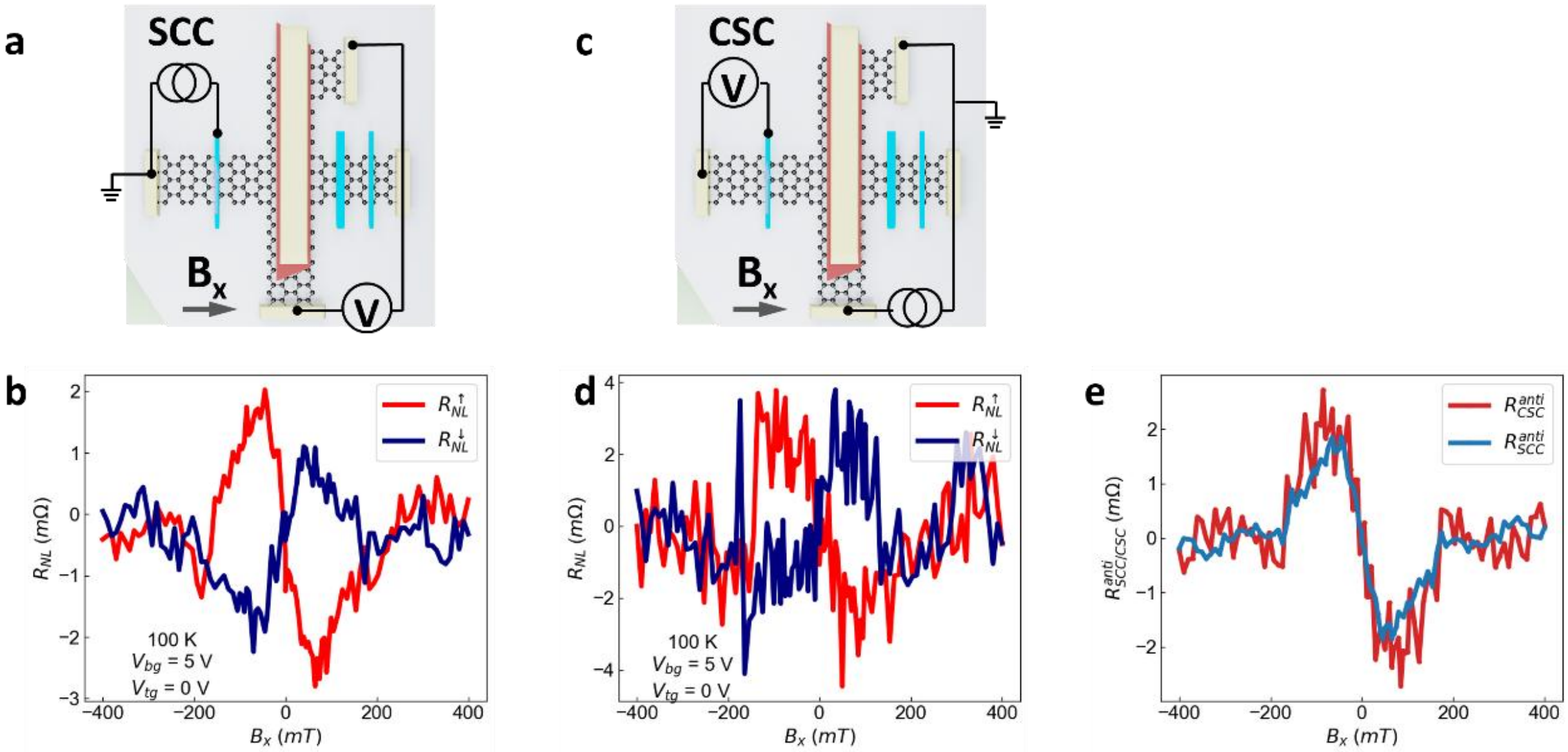

**Fig. S9. Reciprocal spin-to-charge and charge-to-spin (CSC) conversion measurements in Device 2. a,c,** Schematics of the reciprocal measurement configurations. **a,** In the SCC configuration, spins are injected through the ferromagnetic contact and diffuse into the $HfS_2$-covered graphene region, where they are converted into a transverse voltage detected between the normal contacts. **c,** In the CSC configuration, a charge current is applied transversely across the $HfS_2$-covered graphene between the normal contacts, and the generated spin signal is detected nonlocally through the ferromagnetic contact. **b,d,** Nonlocal resistance measured as a function of in the **b,** SCC and **d,** CSC configurations at100 K, $V_{bg} = 5\ V$, and $V_{tg} = 0\ V$. The red and blue curves correspond to opposite magnetization directions of the ferromagnetic contact. **e,** Comparison of the magnetization-dependent, field-antisymmetric SCC and CSC components extracted from c and d. The two signals exhibit comparable amplitudes and the same antisymmetric Hanle-precession profile within experimental uncertainty, consistent with Onsager reciprocity as previously[8] analyzed. The different point-to-point noise levels arise from the different voltage-detection contact configurations.

**Note 8. Control measurements in a dual-gated hBN/graphene device.**

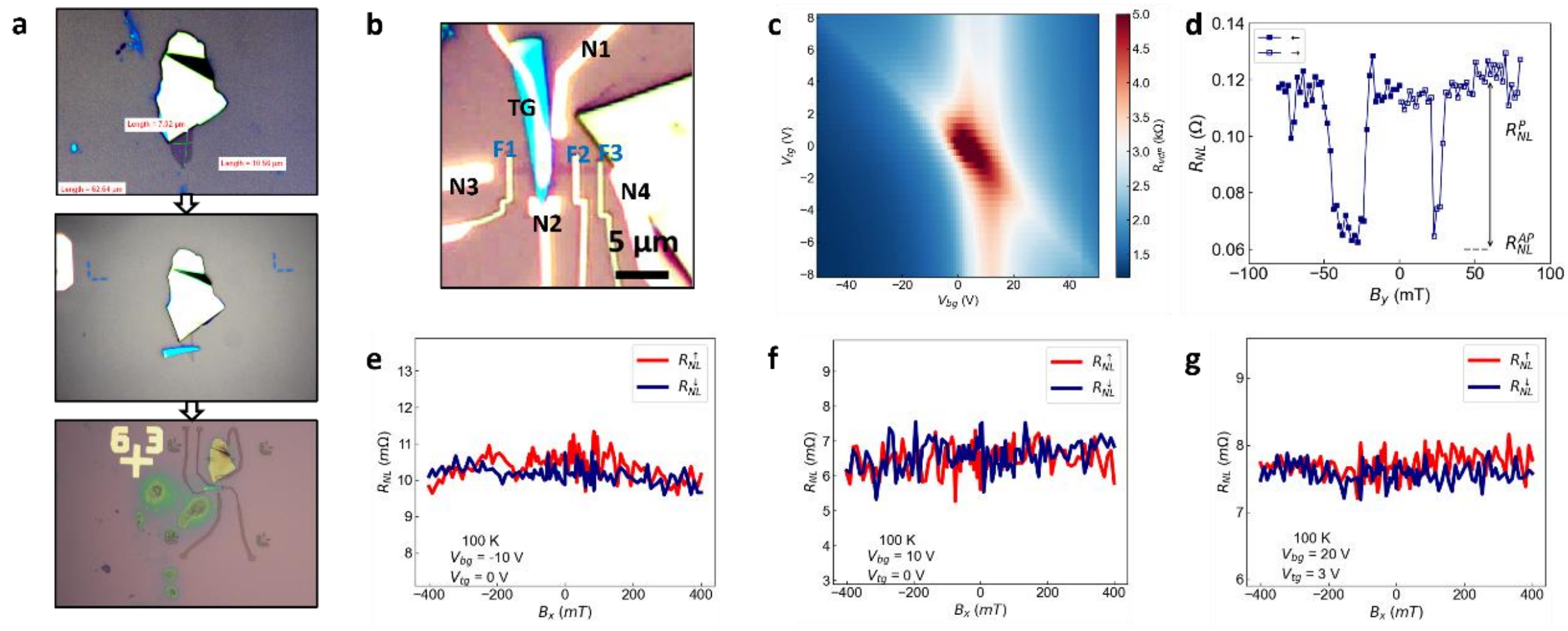

**Fig. S10. Dual-gated hBN/graphene control device and spin–charge conversion measurements. a,** Optical micrographs showing the fabrication sequence of the control device: identification of the trilayer graphene flake on $SiO_2$/Si, dry transfer of an hBN flake onto the graphene, and subsequent channel patterning and electrode fabrication. **b,** Optical micrograph of the completed device. The top-gate electrode is labeled TG, the non-magnetic contacts are labeled $N_1$–$N_4$, and the ferromagnetic contacts are labeled $F_1$–$F_3$. **c,** Dual-gate dependence of the two-terminal resistance measured between $N_1$ and $N_2$. The diagonal high-resistance ridge identifies the CNP trajectory of the top-gated graphene region. **d,** Nonlocal spin-valve signal measured at 300 K and zero gate bias by applying a current between $F_2$ and $N_3$ and detecting the voltage between $F_3$ and $N_4$. Filled and open symbols denote opposite magnetic-field sweep directions. **e–g,** In-plane magnetic field dependent nonlocal resistance measured at 100 K in the SCC geometry, with current applied between $F_2$ and $N_4$ and voltage detected between $N_1$ and $N_2$, at **e,**$V_{bg} = -10\ V$and $V_{tg} = 0\ V$; **f,** $V_{bg} = 10\ V$and $V_{tg} = 0\ V$; and **g,** $V_{bg} = 20\ V$and $V_{tg} = 3\ V$. The red and blue curves correspond to opposite initial injector-magnetization directions. No reproducible magnetization-dependent SCC response is resolved at these representative gate configurations.

**Note 9. Spin precession fitting procedure.**

We model spin diffusion in the symmetric Hanle-precession measurements for parallel and antiparallel FM magnetization configurations, including spin loss at the contacts. The spin propagation in our devices using the Bloch equations:

$$D_s \nabla^2 \vec{\mu} - \frac{\vec{\mu}}{\tau_s} + \vec{\omega} \times \vec{\mu} = 0 \tag{1}$$

where $\vec{\mu}$ is the spin accumulation, $D_s$ the spin diffusion constant, and $\tau_s$ the spin lifetime. $\vec{\omega} = (\omega_x, \omega_y, \omega_z) = g\mu_B\vec{B}$ is the Larmor frequency, $g$=2 is the Landé factor, $\mu_B$ is the Bohr magneton, and $\vec{B}$ is the applied magnetic field.

The spin precession is induced in the $y - z$ plane when applying a magnetic field along $x$ direction. In this case, Eq. (1) turns into

$$D_s \frac{d^2}{dx^2}\begin{pmatrix}\mu_x \\ \mu_y \\ \mu_z\end{pmatrix} - \begin{pmatrix}\tau_x^{-1} \\ \tau_y^{-1} \\ \tau_z^{-1}\end{pmatrix}\begin{pmatrix}\mu_x \\ \mu_y \\ \mu_z\end{pmatrix} + \omega_x \begin{pmatrix}0 \\ -\mu_y \\ -\mu_z\end{pmatrix} = 0 \tag{2}$$

The solution of $\mu_y$ and $\mu_z$ in Eq. (2) are

$$\mu_y = Ae^{\frac{x}{\lambda_s}\sqrt{1+i\omega\tau_s}} + Be^{\frac{x}{\lambda_s}\sqrt{1-i\omega\tau_s}} + Ce^{-\frac{x}{\lambda_s}\sqrt{1+i\omega\tau_s}} + De^{-\frac{x}{\lambda_s}\sqrt{1-i\omega\tau_s}} \tag{3}$$

$$\mu_z = -iAe^{\frac{x}{\lambda_s}\sqrt{1+i\omega\tau_s}} + iBe^{\frac{x}{\lambda_s}\sqrt{1-i\omega\tau_s}} - iCe^{-\frac{x}{\lambda_s}\sqrt{1+i\omega\tau_s}} + iDe^{-\frac{x}{\lambda_s}\sqrt{1-i\omega\tau_s}} \tag{4}$$

where $\lambda_s = \sqrt{\tau_s D_s}$ is the spin diffusion length. $A$, $B$, $C$ and $D$ are the coefficients determined by the boundary conditions. The spin current is defined as:

$$I_{Sy(z)} = -\frac{W_{gr}}{eR_\square}\frac{d\mu_{y(z)}}{dx}, \tag{5}$$

where $W_{gr}$ is the width of the graphene channel and $R_\square$ is the square resistance of graphene. The spin accumulation at the detection position $x = x_{det}$ is then converted into a voltage with

$$V_{det}^s = P\frac{\mu_y(x_{det})}{e}, \tag{6}$$

where $P$ is the spin polarization of the Co detector. We consider that the Co injector has the same spin polarization. $V_{det}^s$ is usually normalized by the charge current $I_c$, giving the non-local resistance

$$R_{NL} = \frac{V_{det}^s}{I_c} = \pm R^s + R^{s0}. \tag{7}$$

$R^{s0}$ is a constant that corresponds to the spin signal at zero magnetic field. The net spin signal $\Delta R_{NL}$ is then expressed as:

$$\Delta R_{NL} = \frac{R_{NL}^P - R_{NL}^{AP}}{2} = R^s. \tag{8}$$

We determine the spin signal using the following boundary conditions: 1. Spin accumulation μ is continuous. 2. Spin current is continuous except: At F1 with spin injection by the charge current, by $\Delta I_s = I_c \cdot P/2$; At F1 and F2 due to the spin backflow effect because of the low contact resistance, by $\Delta I_s = -\mu_s/(2eR_c)$, where $R_c$ is the contact resistance. 3. Spin current is zero at the sample ends.

These boundary conditions in a matrix $X$ which fulfills: $MX = Y$, where $M$ contains the coefficients $M = A, B, C \dots$ and $Y = (0, \dots, I_c P/2, 0, \dots)$ and use the Moore-Penrose inverse to invert $X^{-1}$and obtain $M = YX^{-1}$. Using the numeric solution described above, we input the measured net spin signal $\Delta R_{NL}$, square resistance $R_{\square}$, channel size $W_{gr}$, and the contact resistance $R_c$.

We determine the SHE, REE and UREE efficiency $\theta_{SHE}$ and $\alpha_{REE(UREE)}$, along with the spin lifetime $\tau_s^{TMD/gr}$ of the graphene in the heterostructure as follows.

The non-local resistance $\Delta R_{SHE}$ is given by

$$\Delta R_{SHE} = \frac{\theta_{SHE} R_{\square} \overline{I_{Sz}}}{I_c} \cdot \cos(\beta) \tag{9}$$

where $\overline{I_{Sz}}$ is the average spin current in the cross-junction area of the Hall bar, which is defined as $\overline{I_{Sz}} = \frac{1}{W_{cr}} \int_L^{L+W_{cr}} I_{Sz}(x) dx$. In addition to the spin diffusion term, $\cos(\beta)$ indicates the non-local resistance affected by the contact pulling effect. Here, since the magnetic field is applied in-plane, the magnetization pulling of the FM electrodes is significant because of their finite in-plane shape anisotropy. The magnetization is pulled by an angle $\beta$ from their easy axis towards the field direction, injecting spins along $x$ and resulting in an additional term to the non-local resistance.

$\beta$ is obtained by measuring the non-local resistance of the in-plane symmetrical Hanle precession with two FM. Here $R_{NL}$ obtained in this case is

$$R_{NL} = \pm R^s cos(\beta_1) cos(\beta_2) + R^{s0} sin(\beta_1) sin(\beta_2). \tag{10}$$

For simplicity, we consider that the two Co electrodes have the same pulling effect with the angle $\beta = \beta_1 = \beta_2$. $R^{s0}$ is a constant that corresponds to the spin signal at zero magnetic field. Because of the opposite spin precession sign of $R_{NL}^P$ and $R_{NL}^{AP}$, the sum $R_{sum} = R_{NL}^P + R_{NL}^{AP}$ is then proportional to $\sin^2(\beta)$. For the SCC results for fitting the SHE, we used the contact pulling from the measurement points on the curve of $R_{sum}$ vs $B$.

For the REE and UREE, the non-local resistance $\Delta R_{REE}$ is given by

$$\Delta R_{REE} = \frac{\alpha_{REE} R_{\square} \overline{I_{Sx}}}{I_c} \cos(\beta) \tag{11}$$

where $\overline{I_{Sx}}$ is the average spin current in the cross-junction area of the Hall bar, which is defined as $\overline{I_{Sx}} = \frac{1}{W_{cr}} \int_L^{L+W_{cr}} I_{Sx}(x) dx$, where $L$ is the distance from F1 electrode to the edge of the Hall cross and $W_{cr}$ is the width of the cross. Here, we assume the spin diffusion coefficients for the TMDC covered region and the pristine region to be equal to reduce the number of fitting parameters, and we consider the spin lifetime is isotropic for the proximitized graphene for simplicity. To determine $I_{Sx}(x)$, we use the equations and boundary conditions described above.

Similarly, the non-local resistance $\Delta R_{UREE}$ is given by

$$\Delta R_{UREE} = \frac{\alpha_{UREE} R_{\square} \overline{I_{Sy}}}{I_c} \cos(\beta) \tag{12}$$

where $\overline{I_{Sy}}$ is the average spin current in the cross-junction area of the Hall bar.